\documentclass[12pt]{article}
\usepackage[english]{babel}
\usepackage{graphicx}
\usepackage{verbatim}
\usepackage{epsf}
\usepackage{latexsym}
\usepackage{psfrag}
\usepackage{amsmath,amsfonts,amssymb,amsthm}

\def\bseq{\begin{subequation}}  
\def\eseq{\end{subequation}}
\def\bsea{\begin{subeqnarray}}  
\def\esea{\end{subeqnarray}}

\newcommand{\bbox}{\lower.2ex\hbox{$\Box$}}

\newcommand{\beq}{\begin{equation}}
\newcommand{\eeq}{\end{equation}}
\newcommand{\bea}{\begin{eqnarray}}
\newcommand{\eea}{\end{eqnarray}}
\newcommand{\ena}{\end{eqnarray}}

\newcommand{\Tr}{{\rm Tr}}

\newcommand{\CW}{{\cal W}}

\newcommand{\NN}{{\cal N}}
\newcommand{\ov}{\overline}
\def\beqa{\begin{equation}}
\def\eeqa{\end{equation}}
\def\IR{{\bf R}}
\def\IC{{\bf C}}

\newcommand{\drawsquare}[2]{\hbox{%
\rule{#2pt}{#1pt}\hskip-#2pt
\rule{#1pt}{#2pt}\hskip-#1pt
\rule[#1pt]{#1pt}{#2pt}}\rule[#1pt]{#2pt}{#2pt}\hskip-#2pt
\rule{#2pt}{#1pt}}

\newcommand{\fund}{\raisebox{-.5pt}{\drawsquare{6.5}{0.4}}}
\newcommand{\antifund}{\overline{\fund}}

\newcommand{\be}{\begin{equation}}
\newcommand{\ee}{\end{equation}}

\begin{document}
\setcounter{page}{0}
\begin{titlepage}
\titlepage
\begin{flushright}
SISSA 37/2008/EP\\
CERN-PH-TH/2008-128\\
IFT-UAM/CSIC-08-36\\
\end{flushright}
\vskip 3cm
\begin{center}
\LARGE{\bf E3-brane instantons and baryonic operators\\
  for D3-branes on toric singularities}
\end{center}
\vskip 0.5cm \centerline{{\bf \large Davide Forcella$^{a,b}$ \footnote{\tt davide.forcella@cern.ch}
    I\~naki Garc\'ia-Etxebarria$^{b,c}$ \footnote{\tt inaki@cern.ch}, Angel Uranga$^{b,c}$ \footnote{\tt uranga@mail.cern.ch}}}
\medskip
\footnotesize{

\begin{center}
$^a$ International School for Advanced Studies (SISSA/ISAS) 
\& INFN-Sezione di Trieste, via Beirut 2, I-34014, Trieste, Italy\\
\medskip
$^b$ PH-TH Division, CERN 
CH-1211 Geneva 23, Switzerland\\
\medskip
$^c$ Instituto de F\'{\i}sica Te\'orica UAM/CSIC,\\
Universidad Aut\'onoma de Madrid C-XVI, 
Cantoblanco, 28049 Madrid, Spain
\end{center}}

\bigskip

\begin{abstract}
We consider the couplings induced on the world-volume field theory of D3-branes at local toric Calabi-Yau singularities by euclidean D3-brane (E3-brane) instantons wrapped on (non-compact) holomorphic 4-cycles. These instantons produce insertions of BPS baryonic or mesonic operators of the four-dimensional ${\cal{N}}=1$ quiver gauge theory. We argue that these systems underlie, via the near-horizon limit, the familiar AdS/CFT map between BPS operators and D3-branes wrapped on supersymmetric 3-cycles on the 5d horizon. The relation implies that  there must exist E3-brane instantons with appropriate fermion mode spectrum and couplings, such that their non-perturbative effects on the D3-branes induce operators forming a generating set for all BPS operators of the quiver CFT. We provide a constructive argument for this correspondence, thus supporting the picture.
\end{abstract}

\vfill
\begin{flushleft}
{\today}\\
\end{flushleft}
\end{titlepage}

\newpage

\tableofcontents

\section{Introduction}

The generalization of the AdS/CFT correspondence to dual pairs related
to D3-branes at singularities
\cite{Kachru:1998ys,Klebanov:1998hh,Morrison:1998cs} has provided many
new insights into the duality in situations of reduced supersymmetry
(for instance,
\cite{Bertolini:2004xf,Benvenuti:2004dy,Martelli:2004wu,Martelli:2005tp,Benvenuti:2005ja,Butti:2005sw,Franco:2005sm,Butti:2005vn})
or broken conformal invariance (for instance
\cite{Klebanov:2000hb,Franco:2004jz,Franco:2005zu,Berenstein:2005xa,Bertolini:2005di,Brini:2006ej}). Progress
has been particularly significant for toric Calabi-Yau threefold
singularities, for which there exist powerful tools to study both the
field theory and the CY geometry, like dimer diagrams (aka brane
tilings) \cite{Hanany:2005ve,Franco:2005rj,Hanany:2005ss,Feng:2005gw},
see \cite{Kennaway:2007tq,Yamazaki:2008bt} for reviews. One of the most active topics
in this direction is the identification of gravity duals of the BPS
operators of the CFT and the derivation of BPS operator counting
techniques
\cite{Kinney:2005ej,Biswas:2006tj,Martelli:2006yb,Benvenuti:2006qr,Martelli:2006vh,Herzog:2003dj,Butti:2006au,Hanany:2006uc,Feng:2007ur,Forcella:2007wk,Butti:2007jv,Forcella:2007ps,Forcella:2008bb,Forcella:2008eh}.

BPS operators with low conformal dimension are usually regarded as
dual to supergravity modes \cite{Witten:1998qj}. However the
systematic discussion of general BPS operators, including those with a
number of fields comparable with the number of D3-branes $N$, is most
conveniently carried out by considering all BPS operators to be dual
to systems of supersymmetric D3-branes on the 5d horizon
\cite{Beasley:2002xv}. These are generalizations of the familiar giant
gravitons \cite{McGreevy:2000cw}, and of basic determinant operators
\cite{Gukov:1998kn,Gubser:1998fp,Berenstein:2002ke,Intriligator:2003wr}. Since
most such operators carry non-trivial charges under the baryonic
symmetries of the quiver theory we refer to them as baryonic
operators. Note that in this language mesonic operators are a subset
of baryonic operators having vanishing baryonic charge. The
correspondence between BPS baryonic operators and supersymmetric
wrapped D3-branes has been mostly based on a precise matching of
conformal dimensions and quantum numbers between the two kinds of
objects. Namely, without a more dynamical explanation of the fact that
baryonic operators correspond to wrapped D3-brane states.

\medskip

In this paper we provide a dynamical understanding of the realization
of the gravity dual of baryonic operators in terms of wrapped
D3-branes. Moreover the explanation involves consideration of
euclidean D-brane instantons, concretely E3-branes wrapped on
holomorphic 4-cycles of the CY in the presence of the gauge
D3-branes. In crude terms, the E3-brane instantons leads to insertions
of baryonic operators in the gauge D3-branes, at the level of the
system of D3-branes on the CY geometry. The near horizon version of
the map is that the BPS baryonic operators is related to the boundary
behaviour of the E3-brane, which corresponds to a D3-brane wrapped on
a supersymmetric 3-cycle. The argument is tightly related to the very
suggestive fact \cite{Mikhailov:2000ya}, already exploited in the
literature, that supersymmetric D3-branes on the horizon can be
characterized in terms of holomorphic 4-cycles on the CY
singularity.

\medskip

The holomorphic 4-cycles on which we wrap the E3 instantons are
non-compact, and thus one would say that the instanton action
vanishes. We will assume the existence of some effective cutoff for
the volume of the cycle, generically given by the compactification of
the local geometry we are studying, and we will just be interested on
the prefactor that gives the field theory operator induced by the
instanton, without entering into details of how the setup could be
embedded globally. Which field theory operator is inserted can be
determined by a purely local analysis near the D3 branes.

\medskip

Note that in the above argument, the E3-brane instantons are
considered dynamical, in the sense that its non-perturbative effect is
considered as included in the discussion. This is in contrast with the
recent use of E3-branes on 4-cycles as probes of vevs for baryonic
operators \cite{Klebanov:2007us,Martelli:2008cm}. However there is no
contradiction, but rather a nice agreement, between the two
interpretations of E3-branes on 4-cycles. It is the analog of the
familiar statement \cite{Klebanov:1999tb} that a given AdS field
encodes the information about both the insertion of operators
deforming the CFT, and about the vev of the operator in a given CFT
vacuum dual to a given gravity background. The latter is determined by
the normalizable mode of the AdS field, namely, the component decaying
at the boundary, and can be detected by considering a probe
fluctuation of the field and evaluating its action. Similarly, in
order to measure the vev for a baryonic operator in a given gravity
background, one can introduce a probe with the appropriate
asymptotics, namely given by a D3-branes on a 3-cycle. The
corresponding probe is an E3-brane wrapped a holomorphic 4-cycle on
the CY geometry, and the exponential of its action measures the
vev. This is similar to the computation of a Wilson loop by a
worldsheet with appropriate asymptotics.

\medskip

The relation between E3-brane instanton effects on D3-branes at CY
singularities and BPS operators in AdS/CFT has a direct implication:
the set of BPS operators in the quiver CFT which can be generated from
non-perturbative effects of BPS E3-brane instantons on the CY must
form a generating set of all CFT BPS operators. Also the boundary of a
given E3-brane instanton defines the baryonic D3-brane providing the
gravity dual of the corresponding BPS operator arising from the
non-perturbative effect. Equivalently, the E3-brane on the holomorphic
4-cycle corresponding to the baryonic D3 (i.e. constructed as a cone
over the wrapped horizon 3-cycle) must have a specific structure of
fermion modes charged under the D3-brane theory, and with appropriate
E3-brane world-volume couplings. In this paper we provide a systematic
(and constructive) derivation of this result, for systems related to
D3-branes at toric singularities. This result provides a strong
support for our picture of E3-brane instanton effects as a
first-principle derivation of the AdS/CFT relation between BPS
operators and wrapped D3-branes in AdS/CFT, and of the use of
E3-branes as probes of baryonic vevs.

\medskip

Let us finish this introduction by remarking that the discussion in
this paper is one instance of a very general and deep relation between
instantons in 5 dimensions and baryons, and can be traced back to
early studies of baryons as solitons in the Skyrme model
\cite{Adkins:1983ya}. More recently, this connection has also been
realized in the context of Sakai-Sugimoto models for holographic QCD
\cite{Hata:2007mb}. The results of this paper generalize this
correspondence to the rich class of theories arising from D3 branes at
toric singularities.

\medskip

This paper is organized as follows. In Section \ref{general} we
describe a basic example of the role of E3-brane instantons in systems
of D3-branes in local CY geometries, and its implication for the near
horizon AdS/CFT relation between baryonic operators and wrapped
D3-brane states. In Section \ref{ads/cft} we review the construction
of general BPS operators and their dual wrapped D3-brane states in
AdS/CFT, for systems of branes at singularities. We discuss the
conifold example explicitly, and provide the generalization to
arbitrary toric singularities. In Section \ref{single} we describe the
generation of general BPS 4d field theory operators by E3-brane
instantons, for systems with a single D3-brane. The arguments involve
diverse geometric/field theory operations, such as orbifolding,
partial resolution/Higgsing, as well as a very geometric discussion in
terms of the mirror configuration of E2-brane instantons on systems of
intersecting D6-branes. Our analysis shows a one-to-one map between
field theory BPS operators and 4-cycles on which E3-brane instantons
wrap, which exactly reproduces the AdS/CFT relation. In Section
\ref{narb} we describe the generalization to arbitrary number of
D3-branes, where the map between operators and 4-cycles is more
involved in a sense that we make precise. Finally in Section
\ref{conclu} we present our final comments.

\section{E3-brane instantons and baryonic D3-branes}
\label{general}

Let us consider a configuration of type IIB D3-branes, spanning 4d
Minkowski space $\IR^{1,3}$ and sitting at the singular point of
Calabi-Yau threefold geometry. The gauge theory on the D3-brane
world-volume is determined by the local structure of the singularity
at which the D3-branes sit. We consider the local singularity to be
given by a real cone $C(H)$ over a Sasaki-Einstein 5d manifold $H$.
The low energy dynamics of these branes is a four dimensional
$\mathcal{N}=1$ supersymmetric gauge theory with gauge group $\prod _i
SU(N_i)$, and a set of chiral multiplets in bifundamental
representations, see e.g. \cite{Benvenuti:2005ja,Butti:2005sw,Franco:2005sm,Hanany:2005ss} for
details on the construction of the field theory from geometric data of
the singularity. We adopt the viewpoint that all $U(1)$ factors
(except a decoupled one, which we ignore) are massive due to $B\wedge
F$ couplings with RR 2-forms, and are therefore absent from the
low-energy dynamics. The $N_i$ are positive integers, subject to the
condition of anomaly cancellation or cancellation of localized RR
tadpoles. In this paper we focus on toric singularities, and in the
conformal case $N_i\equiv N$ which automatically satisfies these
constraints.

This type of local systems of D3-branes at CY singularities plays an
important role in two contexts, as local models of type IIB
compactifications to four dimensions, and in the gauge/gravity
correspondence. The latter can be regarded as the near-horizon limit
of the former, leading to derivations of certain results in
AdS/CFT. For instance, the fact that a given AdS field $\phi$ is dual
to certain operator ${\cal{O}}$ in the holographic field theory can be
obtained from the fact that in the original system of D3-branes on CY,
there is a D3-brane world-volume coupling $\phi\, {\cal{O}}$.

In this section we argue that one can draw a similar relation between
baryonic BPS operators in the holographic field theory and AdS
particles from D3-branes on 3-cycles on the horizon, by considering
E3-brane instanton effects on the initial system of D3-branes in a
singular CY geometry.

Let us consider a configuration of D3-branes at a local CY
singularity. It is a natural question to consider the structure of
field theory operators that can be induced by non-perturbative effects
in this setup. There are instanton effects, coming from wrapped
euclidean D-branes
\cite{Becker:1995kb,Witten:1996bn,Harvey:1999as,Witten:1999eg}
(denoted E-branes henceforth) which can induce interesting field
theory operators
\cite{Blumenhagen:2006xt,Ibanez:2006da,Florea:2006si}. In our setup,
BPS instantons preserving half of the 4d $\NN=1$ supersymmetry arise
from E3-branes wrapped on holomorphic 4-cycles in the internal space
\footnote{There are also E$(-1)$ instantons, that we will not
  consider.}. In the non-compact setup, one should distinguish between
E3-branes wrapped on compact or non-compact 4-cycles. The E3 branes
wrapped on compact cycles are classified by the nodes of the quiver,
and correspond to gauge theory instantons when the node is filled by
two or more 4d gauge branes. Even if there is just one or no 4d gauge
branes filling the corresponding node, one can use field theory
techniques to understand the properties of the instanton, see
e.g. \cite{Aharony:2007pr,Argurio:2008jm,Krefl:2008gs}. We focus
instead on E3-branes wrapped on non-compact 4-cycles, passing through
(or near) the singularity, so that they survive in the near horizon
limit to be taken later on. Note that our setup is a generalization of
that recently considered in \cite{Ibanez:2007tu}, with emphasis on a
different motivation.

In the non-compact setup these instantons have vanishing strength, but
such instanton effects become physical when the local model is
embedded in a full-fledged compactification. Some of the properties of
the instanton depend on the global structure of the 4-cycle in the
compactification. For instance, the kind of 4d superspace interaction
they induce is determined by the number of unlifted fermion zero modes
of the instanton. For simplicity, we will assume that the instantons
have only two uncharged fermion zero modes in an appropriate
compactification (the goldstinos of the two 4d $\NN=1$ supersymmetries
they break) and therefore generate a non-perturbative superpotential
(with the measure $d^2\theta$ saturating the two fermion zero
modes). Note that this imposes some specific constraints on the
D3-brane, e.g. to be invariant under the orientifold action on the
compactification, with an $O(1)$ Chan-Paton symmetry. This will not be
very important for our analysis, and in fact the presence of
additional fermion zero modes will simply lead to the insertion of
additional operators ${\ov D}{\ov \Phi}$ in the resulting
multi-fermion F-term, as studied in
\cite{Beasley:2004ys,Beasley:2005iu}, see also
\cite{GarciaEtxebarria:2008pi} for a recent discussion.

Rather, our interest lies in the D3-brane field theory couplings
induced by the non-perturbative instanton effect. The basic structure
of this coupling essentially depends only on the local properties of
the configuration, since it arises from the integration of the charged
fermion modes in the D3-E3 open string sector. These zero modes appear
in the instanton world-volume action via couplings to (combinations of
the) bi-fundamental fields of the 4d field theory, and integration
over them leads to the insertion of a BPS operator of the world-volume
D3-brane field theory. The detailed mapping between E3-branes and BPS
operators will be discussed in coming sections, but it is useful to
present now the basic idea. Consider an E3-brane wrapped on a 4-cycle
passing through the system of D3-branes. The E3-D3 open string sector
leads to charged fermion zero modes $\alpha_{\tilde i}$, $\beta_j$,
where ${\tilde i},j$ are gauge indices. These fields transform as
$\antifund_a$, $\fund_b$, respectively, of the $SU(N)_a\times SU(N)_b$
factor of the D3-brane gauge theory. They couple to a 4d chiral
multiplet $\Phi^{ab}$ in the $(\fund_a,\antifund_b)$ in the instanton
action, as
\begin{equation}
\Delta S_{E3}\, = \, \alpha_{\tilde i}\, \Phi^{ab}_{i{\tilde j}}\,  \beta_j
\end{equation}
The detailed structure of zero modes and the form of the coupling can
be deduced, as we will argue in detail in Section~\ref{single}, from
the cycle wrapped by the instanton and its Chan-Paton
factors. Integrating over the fermion zero modes (and assuming no
extra fermion zero modes beyond the two goldstinos), the instanton
leads to a 4d superpotential
\begin{equation}\label{particular}
\delta_{inst} W \simeq  e^{-T}  \det \Phi^{ab}
\end{equation}
where $T$ denotes the modulus associated to the 4-cycle in an eventual
global embedding of the local configuration, and where the determinant
contracts the color indices, as

\begin{eqnarray}
  \det\Phi^{ab}\, = \, \frac{1}{N!}
  \epsilon_{i_1\cdots i_N}\, \epsilon_{{\tilde j}_1 \cdots {\tilde j}_N}
  \, (\Phi^{ab})_{i_1{\tilde j}_1}\, \ldots\, (\Phi^{ab})_{i_N{\tilde
      j}_N}
\end{eqnarray}

Hence, the above instanton computation leads to a connection between
4-cycles in the singular geometry and BPS (di)baryonic operators in the 4d
field theory. This is an example of the general correspondence to
be studied in Sections \ref{single} and \ref{narb}.

\medskip

Let us now connect the above discussion to the usual AdS/CFT
discussion for baryonic operators. Consider the near horizon limit of
the above system of D3-branes placed on the singularity of $C(H)$. As
discussed in \cite{Klebanov:1998hh,Morrison:1998cs} it corresponds to
type IIB on AdS$_5\times H$, with $N$ units of RR 5-form flux on
$H$. The AdS/CFT implies that this background is exactly equivalent to
the CFT arising from the world-volume D3-brane field theory considered
above.  The precise dictionary relates operators ${\cal{O}}$ of the
CFT to AdS fields $\phi$, in a way that can in many cases be derived
from the existence of a coupling $\phi \, {\cal{O}}$ in the original
D3-brane world-volume field theory. In this sense, it is natural to
expect that the dual of the BPS baryonic operators is related to
E3-branes on CY 4-cycles. In order to make this manifest, recall that
the source for the CFT operator ${\cal O}$ is given by the asymptotic
boundary configuration of the AdS object which produces its
coupling. Thus we may expect that the source for the BPS baryonic
operators is given by the asymptotic boundary configuration of the
E3-brane on the CY holomorphic 4-cycle. The near horizon structure of
a holomorphic 4-cycle is a conical 4-cycle $C(S)$ whose base is a
3-cycle. The state providing the dual to the baryonic BPS operator
$\det \Phi^{ab}$ is thus an AdS particle given by D3-brane wrapped on
the 3-cycle $S$ on the horizon \footnote{By an argument similar to
  \cite{Beasley:2002xv}, we can argue that the asymptotic piece of the
  E3-brane has a Lorentzian continuation to the wrapped D3-brane
  particle.}. This therefore reproduces (and in a sense, explains) the
familiar relation between BPS operators and wrapped D3-branes, and the
relevant role played by holomorphic 4-cycles in their construction
\cite{Beasley:2002xv}, see
\cite{Butti:2006au,Forcella:2007wk,Butti:2007jv}.

\medskip

The above is just an example of a more general correspondence (which
includes BPS mesonic operators as well), which we establish in detail
in this paper. For each BPS operator ${\cal {O}}$ (in a suitable generating set of all BPS operators) in the CFT there exists an E3-brane instanton wrapped on a holomorphic 4-cycle on the
local CY geometry, such that the non-perturbative instanton amplitude
induces an insertion of the operator ${\cal {O}}$ in the D3-brane
world-volume theory. This requires a specific structure of fermion
zero modes and couplings to the CFT fields, which we clarify in
Sections \ref{single} and \ref{narb}.

\medskip

As mentioned in the introduction, the effect of the E3-brane instanton
on the 4-cycle $C(S)$ in the singular CY $C(H)$ leads to an underlying
explanation for two tools which are widely used in AdS/CFT:
\begin{enumerate}
\item The interpretation of a D3-brane wrapping the 3-cycle $S$ on the
  horizon $H$ as the gravity dual of the CFT operator ${\cal {O}}$,
  and thus the general map between BPS operators and supersymmetric
  wrapped D3-branes.
\item The use of E3-brane probes to measure baryonic condensates, since
  these probes provide configurations which asymptote to the baryonic
  D3-brane states in the previous point.
\end{enumerate}

\section{Wrapped branes in AdS/CFT and BPS operators}
\label{ads/cft}

In this section we review the construction of BPS operators in quiver
gauge theories for D3-branes at toric singularities, and the
description of the dual states in AdS/CFT in terms of supersymmetric
D3-branes wrapped on 3-cycles, following \cite{Beasley:2002xv}. The
latter are easily characterized in terms of non-compact 4-cycles of
the singular geometry. We will use the conifold as illustrative
example, but simultaneously discuss the generalization to arbitrary
toric Calabi Yau singularities.

\subsection{Symplectic quotient construction and baryonic charges}

The conifold variety ${\bf X}$ is usually described as the quadric
$xy-wz = 0 $ in $\mathbb{C}^4$, but it can be equivalently described
as a symplectic quotient in the following way. Let us introduce the
four complex variables $x_r$ with $r=1,...,4$. If we give them the
charges $(1,-1,1,-1)$ under a $\mathbb{C}^*$ action we can write the
conifold as the holomorphic quotient
\begin{equation}
{\bf X}=\mathbb{C}^4/(1,-1,1,-1)
\end{equation}
In terms of a symplectic quotient, this corresponds to imposing the
real D-term constraint
\begin{equation}\label{homocon}
|x_1|^2+|x_3|^2-|x_2|^2-|x_4|^2=0
\end{equation}
and quotienting by the $U(1)$ action in the above $\mathbb{C}^*$. To
recover the usual equation for the conifold we consider a basis of the
$\mathbb{C}^*$ invariant monomials $ x = x_1 x_2$, $ y = x_3 x_4$, $ w
= x_1 x_4$, $ z = x_2 x_3$, which satisfy the constraint $xy-wz=0$.

The low energy dynamics of a stack of $N$ D3-branes at the conifold
singularity is a $SU(N)\times SU(N)$ gauge theory with bifundamental
chiral fields $A_1$, $A_2$, and $B_1$, $B_2$ in the
$(\fund,\antifund)$ and $(\antifund,\fund)$ respectively. The chiral
fields interact with the superpotential
$W=\Tr(\epsilon_{ij}\epsilon_{pq}A_iB_pA_jB_q)$. The theory has a
baryonic symmetry under which the fields $A_i$, $B_i$ have charge
$+1$, $-1$, respectively. This baryonic symmetry can be regarded as a
global symmetry arising from a gauge $U(1)$ symmetry in the
$U(N)\times U(N)$ theory, which has acquired a Stuckelberg mass due to
a $BF$ coupling.

The moduli space of the $SU(N)^2$ theory contains the singular
conifold (and all its possible resolutions) in the following way
\cite{Klebanov:1999tb}. Let us restrict ourselves to the $N=1$ case
for simplicity. In this case the gauge group becomes trivial, and the
superpotential vanishes too. The moduli space of such a (free) theory
of 4 complex fields $A_i$, $B_i$ is simply $\mathbb{C}^4$. The Kahler
quotient described above represents the way in which the singular and
resolved conifolds foliate $\mathbb{C}^4$. Imposing the moment map
\begin{equation}
  \label{Dt}
  |A_1|^2+|A_2|^2-|B_1|^2-|B_2|^2=\xi
\end{equation}
selects a particular size for the ${\bf S}^2$ in the base of the
conifold, given by $\xi$. The overall phase of the vevs for the
different fields under the baryonic $U(1)$ encodes the integral of
$C_4$ RR-form over the ${\bf S}^2$.

Notice that there exist a (one-to-one in this case) correspondence
between the homogeneous coordinate $x_r$ in the geometry and the
elementary fields in the gauge theory $A_i$, $B_j$. In particular the
$\mathbb{C}^*$ action of the symplectic quotient construction is just
the complexification of the baryonic symmetry in gauge theory.  This
is just a reflection of the familiar statement that the mesonic moduli
space of a D3-brane is the transverse geometry, see
\cite{Forcella:2008bb,Forcella:2008eh} for a recent discussion of the
mesonic and baryonic moduli spaces of D3-branes at singularities.

\medskip

The above structure generalizes to arbitrary toric singularities. This
follows from their definition as symplectic/holomorphic quotients of
${\mathbb C}^d$ by an abelian group $K\sim (\mathbb{C}^*)^{d-3}\times
\Delta$, where $\Delta$ is a discrete group. Indeed, like in the
conifold case, there is a relation between homogeneous coordinates in
the symplectic quotient construction and chiral multiplets of the
D3-brane gauge theory. The relation however is in general not
one-to-one, and to each homogeneous coordinate in the geometry is
associated more than one chiral superfield \cite{Butti:2005vn}. Also
the D3-brane field theories have a set of baryonic symmetries, which
can be regarded as the $U(1)$ factors in the $U(N)\times \ldots\times
U(N)$ theory, eventually massive by the Stuckelberg mechanism. In
analogy with the conifold case, these baryonic symmetries can be
related to the $U(1)$ symmetries in the symplectic quotient
construction \cite{Forcella:2008bb,Forcella:2008eh}.

\subsection{The general set of BPS operators}\label{genbpsop}

According to the AdS/CFT correspondence the low energy gauge theory of
$N$ D3-branes on $\mathbb{R}^{1,3}\times C(H)$ is dual to string
theory on the background $AdS_5 \times H$, for a general CY conical
singularity $C(H)$ with base a 5d Sasaki-Einstein compact manifold $H$
\cite{Klebanov:1998hh,Morrison:1998cs}. In particular, the AdS/CFT
correspondence predicts a one-to-one map between the BPS gauge
invariant operators on the field theory side and the BPS states on the
gravity side.

Let us review this correspondence for the case of the conifold ${\bf
  X}=C(T^{1,1})$, whose gauge theory is dual to string theory on
$AdS_5 \times T^{1,1}$.  For our purposes it is useful to start by
considering the simplest baryonic operators $\det A_i$, $\det B_j$.
\begin{eqnarray}
\label{exbar}
\epsilon_{p_1,...,p_N}  \epsilon^{k_1,...,k_N} (A_{i_1})^{p_1}_{k_1}... (A_{i_N})^{p_N}_{k_N} = (\det A)_{ ( i_1,...,i_N ) } \nonumber \\
\epsilon_{p_1,...,p_N}  \epsilon^{k_1,...,k_N} (B_{i_1})^{p_1}_{k_1}... (B_{i_N})^{p_N}_{k_N} = (\det B)_{ ( i_1,...,i_N ) }
\end{eqnarray}
As has been studied in \cite{Gubser:1998fp}, the AdS states
corresponding to these BPS operators are static D3-branes wrapping the
${\bf S}^3$ contained in the horizon manifold $T^{1,1}$ (with a
specific orientation). The specific 3-spheres are easily described
using the homogeneous coordinates. Given a supersymmetric 3-cycle
$C_3$ on the horizon manifold, the real cone $C(C_3)$ over it defines
a holomorphic non-compact 4-cycle on the Calabi-Yau singular geometry,
which can be described as the zero locus of the homogeneous
coordinates. The baryonic operators $\det A_i$, $\det B_j$ correspond
to the 4-cycles $x_r=0$.

This basic idea can be exploited to reproduce the full spectrum of BPS
operators of the conifold theory, which includes many other
operators. Indeed, the above are just the baryonic operators with the
smallest possible dimension: $\Delta _{\det A ,\det B} = N \Delta
_{A,B}$.  The full set of BPS operators with the same baryonic charges
as e.g. $\det A$ can be constructed as follows. Following
\cite{Beasley:2002xv,Berenstein:2002ke} (with a different notation) we
define the operators
\begin{equation}
\label{AAA}
{\cal A}_{\cal P}= A_{i_1}B_{j_1}...A_{i_m}B_{j_m}A_{i_{m+1}}
\end{equation}
Namely, we construct an operator in the $(\fund,\antifund)$ with the
same gauge and baryonic charges as $A$, by concatenating a number of
bifundamental fields with indices contracted, in a pattern encoded in
the multi-index ${\cal P}$. In terms of the quiver we associate an
operator to any path, which we also denote ${\cal P}$, obtained by
concatenation of arrows corresponding to $A$- and $B$-fields.

Given a set of $N$ (possibly different) operators of that kind,
denoted ${\cal A}_{{\cal{P}}_1},\ldots, {\cal A}_{{\cal{P}}_N}$, we
can construct the general `$A$-type' baryonic operator as
\begin{equation}
\label{genbarA}
{{\cal{O}}^A}_{\{ {\cal P} \} } \,=\, \epsilon_{{\tilde p}_1,\ldots,{\tilde p}_N} \, \epsilon_{k_1,\ldots,k_N} ({\cal A}_{{\cal {P}}_1})_{p_1{\tilde k}_1}\, \ldots\, ({\cal A}_{{\cal {P}}_N})_{p_N\tilde{k}_N} 
\end{equation}
One can similarly define $B$-type operators. Operators (\ref{genbarA})
provide the generalization of the simplest baryonic operators
(\ref{exbar}). Note that e.g. all $A$-type operators carry the same
baryonic charges, but are of different conformal dimension.  This set
of operators provides a basis of all BPS operators in the gauge theory
(with mesonic operators arising from products of $A$- and $B$-type
operators, so that they carry no baryonic charge, and baryonic
operators of higher or lower baryonic charge coming from products of
$A$-type and $B$-type operators respectively).

\medskip

It is possible to generalize this discussion to general toric
singularities \footnote{For studying gauge theories dual to D3 branes
  at toric singularities it is most convenient to use dimer model
  techniques, which also play an important role in our subsequent
  analysis. We include for convenience a short introduction to dimer
  models in Appendix~\ref{dimerreview}.} as follows
\cite{Butti:2006au}. Given one bifundamental chiral multiplet
$\Phi_{a,b}$ in the $(\fund_a,\antifund_b)$, one can form the basic
di-baryonic operator generalizing (\ref{exbar}) by taking its
determinant $\det(\Phi_{a,b})$. This corresponds to the BPS operator
with lowest dimension in the corresponding sector of baryonic
charges. More in general, one can construct an operator with baryonic
charges proportional to $N$ under the baryonic symmetries $U(1)_{a,b}$
(not necessarily connected by a single arrow) by considering $N$
(possibly different) paths ${\cal {P}}_1,\ldots{\cal {P}}_N$, in the
quiver, joining the nodes $a,b$ \footnote{Since the operators are
  defined modulo F-terms, it is more practical to define the operator
  using paths joining faces in the dimer diagram. The equivalence
  modulo F-terms is related to the equivalence of paths under homotopy
  deformations. Hence different paths correspond to homotopically
  different paths between the faces $a,b$.}. Using the corresponding
operators ${\cal {O}}_{{\cal{P}}_1}, \ldots, {\cal{O}}_{{\cal{P}}_N}$,
all of which transform in the $(\fund_a,\antifund_b)$, we can
construct
\begin{equation}
  \label{OPath}
  {\cal O}_{\{P\}}\, =\, \epsilon_{\tilde{p}_1,\ldots,\tilde{p}_N} \, \epsilon_{k_1,\ldots,k_N} \, ({\cal {O}}_{{\cal{P}}_1})_{p_1\tilde{k}_1}... ({\cal {O}}_{{\cal {P}}_N})_{p_N\tilde{k}_N}\, . 
\end{equation}
Observe that, as in the conifold case, once we increase the baryonic
charges we are interested in, we are forced to consider product of
operators like (\ref{OPath}).

\subsection{The gravity duals and holomorphic 4-cycles}
\label{gravity-duals}

The above description is well-suited to provide a construction of the
states dual to these BPS operators. Going back to the conifold
example, recall that the basic baryonic operators (\ref{exbar}) are
mapped to static D3-branes wrapping specific three cycles of $T^{1,1}$
in a volume-minimizing fashion. Since we would like to describe states
dual to operators with the same baryonic charge but higher conformal
dimension, we need to describe supersymmetric D3-branes wrapped on the
same homology class, but not in a volume-minimizing fashion. The state
nevertheless manages to remain BPS due to a non-trivial motion in the
horizon geometry, as for the giant gravitons in
\cite{McGreevy:2000cw}.

These states once again have a nice correspondence with holomorphic
divisors on the singular Calabi-Yau geometry. Recall that the baryonic
charge of the simplest baryonic states (\ref{exbar}) is related to the
$\mathbb{C}^*$ charge of the function whose zero locus defines the
4-cycle, namely $x_i$. Hence, the BPS operators in the same baryonic
charge sector, but with higher conformal dimension, are expected to
correspond to 4-cycles defined as the zero locus of a more general
function of holomorphic coordinates, with the same degree of
homogeneity under the $\mathbb{C}^*$ action. More formally, they
correspond to different sections of the same non-trivial line bundle
over the CY variety.

Consider for example the case of a single D3-brane $N=1$, and the set
of 4-cycles corresponding to BPS operators with baryonic charge
$B=1$. This is
\begin{eqnarray}\label{sec}
f_{B=1}(x_1,x_2,x_3,x_4)&=&c_{1}x_1+ c_{3}x_3 + \nonumber\\
& & c_{11;2}x_1^2 x_2 + c_{13;2} x_1 x_3 x_2 +  c_{33;2}  x_3^2 x_2 + \nonumber\\
& &  c_{11;4}x_1^2 x_4 + c_{13;4} x_1 x_3 x_4 +  c_{33;4}  x_3^2 x_4 + ....
\end{eqnarray}  
where the coefficients, collectively denoted $c_P$, parametrize the
complex structure of the divisor. This infinite family of holomorphic
4-cycles provides a description of all the possible supersymmetric
D3-branes wrapping the ${\bf S}^3$ in $T^{1,1}$ with positive
orientation. The space parametrized by the $c_P$ is a classical
configuration space for the particles arising from the D3-brane, which
has to be properly quantized. Namely, the gauge theory BPS operators
should correspond to appropriate wavefunctions on the space
parametrized by $c_P$. Using geometric quantization, one can determine
that the different wavefunctions are given by degree-$N$ monomials on
the $c_P$ \cite{Beasley:2002xv}. We denote $|c_{P_1},\ldots,
c_{P_N}\rangle$ the state corresponding to the wavefunction \beqa
\Psi(\{ c_P\})=c_{P_1}\ldots c_{P_N}.
\label{wavefunction}
\eeqa This state defines a particle in $AdS_5$, whose dual BPS
operator is obtained as follows: using the relation between monomials
in $x_r$ and bi-fundamental fields $A,B$, the monomial corresponding
to each $c_{P_\alpha}$ corresponds to an operator
${\cal{A}}_{P_\alpha}$ of the form (\ref{AAA}), or its B-type
analog. The BPS operator dual to the state $|c_{P_1},\ldots,
c_{P_N}\rangle$ is given by the operator ${\cal {O}}_{ \{ P \}}$
defined in (\ref{OPath}). More general BPS operators can be generated
by taking products of these.

The states in AdS side $|c_{P_1},\ldots,c_{P_N}\rangle$ correspond to
wavefunctions related to a set of $N$ (coefficients of) such monomials
in the homogeneous coordinates of $C(H)$.  The corresponding BPS
operator is a baryonic operator given by (\ref{OPath}), or suitable
products thereof.

\medskip

This procedure extends to generic toric singularities
\cite{Butti:2006au}. For a general toric singularity there is also a
correspondence between a monomial in the homogeneous coordinates
(hence its coefficient $c_P$ in a general expansion) and operators
(denoted ${\cal O}_{\cal{P}}$) given by a product of bifundamental
fields describing a path ${\cal{P}}$ in the quiver/dimer diagram of
the gauge theory. The major difference with the conifold case is that
in the generic case the correspondence between the homogeneous
coordinate and the fields is one to many, as studied in detail in
\cite{Butti:2007jv}. The issue here is that, if the 4-cycle wrapped by
the instanton has a non-trivial homotopy group, we can construct
different nontrivial flat bundles on the 4-cycle, and this information
about the bundle must be specified together with the purely
geometrical data in order to completely determine the map between
wrapped branes and BPS operators. This makes passing from the case of
the conifold to the case of the general toric singularity very
nontrivial. In fact, to our knowledge, only in the orbifold case
(which we discuss in detail in Section~\ref{long-baryonic-couplings})
is this map well understood in terms of the explicit data of the
divisor and the bundle \footnote{We thank the referee for emphasizing
  this point to us.}.

Nevertheless, in \cite{Butti:2007jv} a generic method to compute the
multiplicities of the map from cycles to operators is proposed, and it
agrees well with the field theory result for nontrivial toric
singularities. This method admits a nice interpretation in the
manifold mirror to the toric variety. In Appendix~B of
\cite{Butti:2007jv} it is discussed how once one goes to the mirror
type IIA side, the extra bundle data gets encoded into topological
information of the cycles wrapping the mirror surface $\Sigma$ (for
convenience, we have included a short review of the relevant concepts
in Appendix~\ref{mirror}). We will use similar ideas in
Section~\ref{mirror-picture} in order to give evidence for our results
in the case of geometries with multiplicities, which are less
understood from the type IIB side.

\medskip

Thus, the AdS/CFT correspondence between BPS operators and wrapped
D3-branes is based on associating a holomorphic 4-cycle in the CY
singularity to each concatenated chain of bi-fundamentals in the field
theory, in a way determined by the relation between homogeneous
coordinates (plus information about the bundle) and bifundamental
fields. Our proposal to provide a first principle derivation of this
AdS/CFT map requires that the E3-brane instanton wrapped on the
4-cycle induces a non-perturbative insertion of precisely the dual BPS
operator on the D3-brane field theory. This is explicitly shown for
toric singularities in the next two sections, by a combination of
techniques.

\section{BPS operators from E3-brane instantons: \\ The single D3-brane case}
\label{single}

In this section we consider E3-brane instantons on non-compact
holomorphic 4-cycles in general toric CY geometries, in the presence
of a single D3-brane. We argue that they provide a correspondence
between 4-cycles in the singular geometry and BPS operators
corresponding to (part of the) 4d effective operator induced by the
instanton. This correspondence is in fact nicely correlated with the
one described in the previous section, lending support to our
identification of E3 instantons with baryonic operators.

\subsection{General considerations and result}
\label{generalities}

Before going into details, let us summarize here the result we want to
show, and the strategy that we will follow in order to show it.

\medskip

In this section we will restrict the discussion to the $N=1$ case
(here $N$ denotes the number of branes in the singularity), which
already allows us to discuss the precise form of the one-to-one map
between BPS operators (and their wrapped D3-brane duals) and E3-brane
instanton effects on D3-branes on the CY. We postpone the discussion
of the complications arising from having $N$ arbitrary to
Section~\ref{narb}. Although we do not provide a formal proof, we
present a sufficiently general line of argument, illustrated in
several explicit examples. Also, notice that the sugra approximation
is expected to break down for the $N=1$ case, since the background
will become strongly curved. Nevertheless, we expect supersymmetry to
protect the BPS sector and allow the discussion in terms of
holomorphic curves. Also, as we will discuss in Section~\ref{narb},
the lessons we learn from studying this simple case in the geometric
regime can be carried over easily to the regimes of larger $N$, where
the sugra approximation is well justified.

Let us start by stating our general proposal. Since for $N=1$ the
gauge group is trivial, the set of ``single determinant'' BPS
operators \footnote{The name ``single determinant'' comes from the
  fact that in the case of general $N$ such concatenated chains give
  operators that can be written as a single determinant of the chain
  of fields. As we will discuss in Section~\ref{narb}, the set of
  single-determinant operators generate the whole set of BPS
  operators.} is described as the set of concatenated chains of
bi-fundamental multiplets, modulo F-terms. Equivalently, operators
carrying baryonic charges $\pm 1$ under two baryonic symmetries $a,b$,
are associated to paths ${\cal{P}}_\alpha$ joining the corresponding
faces in the dimer diagram, modulo homotopy transformations (see
appendix~\ref{dimerreview} for a short review on dimer diagrams). We
denote this operator by ${\cal O}_{{\cal{P}}_\alpha}$. Note that the
indices $a,b$ are implicit in this notation, and that we also use it
for mesonic operators, for which the paths are closed loops in the
dimer.  Let us denote $\Sigma_{{\cal{P}}_\alpha}$ the 4-cycle that
corresponds to one such operator by the AdS/CFT correspondence
\cite{Butti:2006au}, as described in the previous section. In this
section we argue that, considering the configuration of a single
D3-brane at the CY singularity, the operator ${\cal
  O}_{{\cal{P}}_\alpha}$ is precisely generated as (part of) the
amplitude of an E3-brane instanton wrapped on
$\Sigma_{{\cal{P}}_{\alpha}}$.

The appearance of ${\cal O}_{{\cal{P}}_\alpha}$ in the instanton
amplitude can be regarded as arising from the integration over fermion
modes in the E3-D3 open string sector, $\beta$, $\gamma$, in the
$\antifund_a$, $\fund_b$, respectively, with a coupling in the
instanton world-volume action
\begin{equation}
  \beta \, {\cal
    {O}}_{{\cal{P}}_{\alpha}}\, \gamma.
\label{longc}
\end{equation}
For mesonic operators, the modes $\beta$, $\gamma$ form a vector-like
pair. When ${\mathcal{O}_{\mathcal{P}}}$ involves several
bifundamental chiral multiplets, we refer to these couplings as
``long''. The operator $\mathcal{O}_{\mathcal{P}}$ takes
zero vacuum expectation value exactly on the four cycle
$\Sigma_{{\cal{P}}_{\alpha}}$, while it gives mass to the modes
$\beta$, $\gamma$ away from $\Sigma_{{\cal{P}}_{\alpha}}$. This fact
is a consistency check that the coupling (\ref{longc}) is generated by
an E3-brane instanton wrapped on $\Sigma_{{\cal{P}}_{\alpha}}$.

Notice that the complete structure of the instanton amplitude may
contain additional insertions, due to extra fermion modes, etc, which
actually depend on the details of the global compactification. As
explained in the introduction, we center our analysis in this part of
the instanton prefactor, which depends only on local properties of the
configuration.

\medskip

The simplest case in which we claim that our proposal holds is the
case of a E3-brane wrapped in a single irreducible cycle, which we
expect to be associated to an operator ${\cal{O}_{\cal {P}}}$ which
does not factorize. We expect this close relation between
factorizability of the cycle and the operator to hold in
general. Nevertheless, this is a somewhat subtle point, and we want to
clarify it in the following.

As discussed at the end of Section~\ref{gravity-duals}, in the case of
a general toric singularity it is important to include the bundle data
in the specification of the string dual to the baryon operator. When
we speak of factorizability and recombinability here, it is understood
that the bundle should be taken into account. More simply, one could
frame the discussion in the mirror manifold, as we will do in
Section~\ref{mirror-picture}.

Another issue is that, since in fact for $N>1$ any BPS operator can be
factorized as a product of bifundamentals, we should clarify what
happens for cycles which are reducible but can be recombined into one
smooth irreducible cycle \footnote{A related issue for the case $N>1$
  is that any operator of the form $\det AB$ factorizes as $\det A
  \det B$.}. When $\Sigma_{P_\alpha}$ is reducible, our map implies
that the corresponding operator is generated by a multi-instanton
process, with one E3-brane wrapped on each component of the reducible
4-cycle (see \cite{Blumenhagen:2007bn} for instantons on reducible
cycles, and
\cite{GarciaEtxebarria:2008pi,GarciaEtxebarria:2007zv,Blumenhagen:2008ji,Cvetic:2008ws}
for recent papers on multi-instantons). Multi-instantons imply
additional zero modes, and the discussion of their 4d amplitude is
more involved. Nevertheless, we argue that the general statement of
the relation between 4-cycles $\Sigma_{\cal {P}}$ and BPS operators
${\cal {O}}_{\cal {P}}$ holds in general, by applying the following
deformation argument. It is possible to regard the reducible cycles
$f=0$, e.g. $xy=0$ or $x^2=0$, as singular limits of irreducible
4-cycles $f+\epsilon g=0$ like $xy-\epsilon=0$ or $x^2-\epsilon
y=0$. Note that here $\epsilon$ should be regarded not as an instanton
bosonic mode, to be integrated over, but rather as a tunable parameter
fixed by boundary conditions, or the complex structure moduli of the
global compactification \footnote{We force the recombination of the
  instantons by changing their complex structure. See
  \cite{Blumenhagen:2007bn} for a discussion of instanton
  recombination by motion over Kahler moduli space.}. The 4d amplitude
of the irreducible instanton leads to the 4d operator $f+\epsilon g$,
as in the above paragraphs. In the limit $\epsilon\to 0$, the
instanton becomes reducible and seemingly more complicated. However,
the dependence in $\epsilon$ determined away from that point can be
extended using holomorphy \footnote{See
  \cite{GarciaEtxebarria:2008pi,GarciaEtxebarria:2007zv} for a general
  discussion of holomorphy of non-perturbative superpotential and
  higher F-terms, and reducible instantons in loci in Kahler moduli
  space. Although here we are interested in the (much more
  holomorphic) discontinuity complex codimension one loci in the
  complex structure moduli space in the spectrum of BPS branes, the
  microscopic analysis for those systems could be carried out in a
  similar spirit for the systems at hand to show the continuity of the
  4d contribution as $\epsilon\to 0$.} of the 4d non-perturbative
F-term in the (complex structure dependent) parameter
$\epsilon$. Hence at $\epsilon=0$ it must reduce to just an insertion
of $f$ (despite the fact that the process generating this insertions
may be rather involved).

Thus we expect our general arguments to apply even for reducible
4-cycles which admit a recombination into a single smooth one,
$\Sigma_{\cal{P}}={\Sigma_1}+\ldots+\Sigma_K$. The complete operator
generated by the multi-instanton process defined by $\Sigma_{\cal
  {P}}$ is given by the concatenation of the operators generated by
the different instantons associated to the individual
$\Sigma_i$. Clearly this implies that all ``single determinant''
operators of the $N=1$ theory, defined by a path of concatenated
bifundamentals, can be regarded as generated by a recombinable
multi-instanton of this kind. On the other hand, ``multi determinant''
operators of the $N=1$ theory, defined by products of the above,
namely by several non-concatenable paths, correspond to
multi-instanton processes which do not admit a recombination into a
single smooth one. Correspondingly, these are indeed described by
multi-particle states in AdS/CFT, arising from different wrapped
D3-branes. We therefore focus on ``single determinant'' operators,
since they contain all the essential information about the spectrum of
BPS operators. We will apply the above considerations about
recombinability when necessary, and even abuse language using
couplings like (\ref{longc}) in such situations, and treating the
process as a single-instanton one.

\medskip

As mentioned above we will not provide a formal proof of the
correspondence, but we will argue in several different ways for the
existence of the couplings and zero modes that we require. Let us
provide here a short summary of the arguments in the rest of this
section.

Sections \ref{sec:single-field} and \ref{mesonic} review some results
already known in the literature which support our viewpoint, for the
particular cases of single field and mesonic operator
insertions. Section~\ref{long-baryonic-couplings} argues that
couplings of the form~(\ref{longc}) are present for any orbifold
singularity. The argument proceeds essentially by orbifolding the
results known from flat space. Then, by using partial resolution,
Section~\ref{long-general} argues that such couplings are present for
any toric singularity.

Section~\ref{mirror-picture} gives an independent argument for the
validity of our result. We show explicitly how we can find the disc
worldsheet instantons giving the coupling~(\ref{longc}) in some simple
situations. This picture has the advantage that everything is
geometrical (in particular, there are no subtleties having to do with
Chan-Paton factors), but also has the drawback that the special
Lagrangian cycles dual to the cycles wrapped by the instanton are not
explicitly know. Nevertheless, we argue that with a reasonable ansatz
for the topology of the dual cycles (based on the well-understood
single insertion case), one obtains couplings of the
form~(\ref{longc}).

\subsection{Single field insertions}
\label{sec:single-field}

The simplest BPS operators of the form described above in the $N=1$
theory are given by the bi-fundamental chiral multiplets $\Phi_{ab}$
themselves. The correspondence between branes wrapped on 4-cycles and
such bifundamentals has already been considered in appendices of
\cite{Franco:2006es} for D7-branes, and of \cite{Ibanez:2007tu} for
E3-brane instantons. Indeed, using dimer diagrams it is
straightforward to verify that to a given bi-fundamental multiplet one
can associate a divisor in the singular geometry, such that an
E3-brane wrapped in the latter has fermion zero modes $\beta$,
$\gamma$, coupling as $\beta \hbox{ }\Phi_{ab} \gamma$. In our present
setup, we regard this result as the simplest realization of the
correspondence between 4-cycles and BPS operators of the $N=1$
theory. In fact, it was already argued in these papers that this
correspondence is exactly the same as that obtained from the AdS/CFT
correspondence.

\subsection{Mesonic operators}
\label{mesonic}

Let us argue that this correspondence applies also to mesonic
operators, a discussion in fact related to systems studied in
\cite{Ganor:1996pe,Baumann:2006th,Berg:2004ek}. The consideration of
mesonic operators will naturally provide us with examples of the
correspondence beyond single field insertions.

Consider the simplest situation of a single D3-brane in flat
transverse space $\mathbb{C}^3$, parametrized by
$(z_1,z_2,z_3)$. Abusing notation, we also denote $z_i$ the D3-brane
adjoint chiral multiplets, parametrizing the D3-brane position.
Consider an E3-brane instanton wrapped on the 4-cycle defined by
e.g. $z_1=0$. In the E3-D3 open string sector there are fermion modes
$\beta,\gamma$, with a world-volume coupling $\beta z_1\gamma$, which
reflects that the separation of the branes in $z_1$ controls the mass
of these modes. Thus, integration over these instanton fermion modes
leads to an insertion of the mesonic operator $z_1$, similarly to the
previous section. Notice that we manifestly recover the AdS/CFT map
between the 4-cycle $z_1=0$ and the BPS operator $z_1$.

This is just the E3-brane version of the result in \cite{Ganor:1996pe}
for non-perturbative effects of D7-branes in presence of D3-branes. It
is also a particularly simple realization of the effect computed in
\cite{Baumann:2006th} \footnote{Related results appear in
  \cite{Berg:2004ek}, where the computations are done from the open
  string viewpoint. This is similar in spirit to our computations in
  this section.}. In this paper, the authors considered the
non-perturbative superpotential generated by gaugino condensation on
D7-branes wrapped on a non-compact 4-cycles, in a warped deformed
conifold background, as a function of the location $z_i$ of one
D3-brane. The result involved a computation of the change of warped
wrapped volume as a function of this position, leading to a
modification of the instanton amplitude of the form (adapting already
to E3-brane instantons rather than the fractional instantons involved
in D7-brane gaugino condensates),
\begin{equation}
  S\, =\,\int \, d^2\theta\,
  f(z_i)\, e^{-T}
\label{baumann}
\end{equation}
where $f(z_i)=0$ is the equation of the 4-cycle wrapped by the
instanton brane.

In fact, many of the ingredients of the configuration, like the 3-form
fluxes, the complex deformation, or even the fact of being at a
conifold, are actually not essential. The result has much more general
validity, since it amounts to a computation in the closed string
channel of the annulus diagram that corresponds to integrating over
E3-D3 instanton fermion modes. Applied to our flat space example, the
instanton wrapped on the 4-cycle $z_1=0$ leads to the insertion of the
(mesonic) operator $z_1$.

The argument applies to general singularities. Since a general mesonic
operator correspond to a holomorphic function $f(z_i)$ on the singular
geometry, an instanton wrapped on the divisor $f(z_i)=0$ leads to a 4d
effective vertex containing the mesonic operator $f(z_i)$. From the
viewpoint of the instanton, this arises from integrating over E3-D3
fermion modes $\beta$, $\gamma$, with couplings $\beta f(z_i)\gamma$,
reflecting that they become massive as the E3-brane is moved away from
the D3-branes (namely, when the defining equation is modified to
$f(z_i)=\epsilon$).  This shows the existence of general ``long''
couplings of the form (\ref{longc}), for mesonic operators.

For example consider a single D3-brane on a conifold described as $xy-zw=0$, and an E3-brane wrapped on $z=0$. In terms of the underlying D3-brane field theory, the coordinates are mesonic operators, 
\begin{eqnarray}
A_1B_1=x \hbox{ , } A_2B_2=y \hbox{ , } A_1B_2=w \hbox{ , } A_2B_1=z 
\end{eqnarray}
So the non-perturbative E3-brane instanton reads (assuming it generates a superpotential)
\beqa
\int\, d^2\theta\, A_1 B_2\, e^{-T}
\eeqa
Hence in general, for any given mesonic operator ${\cal{O}}_{\cal{P}}$ of the $N=1$ theory there is a 4-cycle such that the wrapped E3-brane instanton leads to an insertion of ${\cal {O}}_{\cal{P}}$ in the 4d effective action.

Notice the fact that the couplings of the form (\ref{longc}) involve
the operator ${\cal O}_{\cal{P}}$ modulo F-terms should be clear at
this point. In fact, the rewriting of a mesonic operator in terms of
the underlying fields is an operation which is defined modulo the
F-term relations.

\subsection{Long baryonic couplings for orbifolds}
\label{long-baryonic-couplings}

We have argued that instantons can generate a variety of long
couplings and BPS operators for some simple singularities. One simple
way to show the appearance of long couplings in more general and more
involved singularities is orbifolding. For instance, we may consider
orbifolds of $\mathbb{C}^3$ by a discrete subgroup $\Gamma$ of
$SU(3)$, which we take to be abelian in order for the orbifold to
admit a toric description. The gauge group splits as a product
(maintaining the $U(1)$ for momentary convenience) of $U(1)^K$ with
$K$ the order of $\Gamma$, and each adjoint of the parent theory leads
to a set of bi-fundamentals. For instance, considering the
$\mathbb{Z}_3$ orbifold generated by a rotation $(\frac 13,\frac
13,-\frac 23)$, the three adjoints $X$, $Y$, $Z$ lead to bifundamental
fields $X_{i,i+1}$, $Y_{i,i+1}$, $Z_{i,i+1}$, transforming in the
$(\fund_i,\antifund_{i+1})$.  The superpotential of the theory is
obtained by replacing the original adjoints in the parent
superpotential by the bifundamentals they lead to, in all possible
ways consistent with gauge invariance. Namely \beqa
\label{supoz3}
W\, =\, X_{i,i+1} Y_{i+1,i+2} Z_{i+2,i}\, -\, X_{i,i+1} Z_{i+1,i+2} Y_{i+2,i}
\eeqa

In this orbifolding process, the fate of E3-brane instantons is easy
to determine.  In performing the quotient, one needs to specify the
action of $\Gamma$ on the Chan-Paton indices of open strings with
endpoints on the E3-brane. The choice of this Chan-Paton phase,
determines with which of the possible bi-fundamentals the
corresponding E3-D3 fermion zero modes will couple in the quotient
theory.

Consider a concrete example, corresponding to instantons leading to
single field insertions. Consider an E3-brane defined by $X=0$ in the
$\IC^3$ theory, thus leading to the insertion of the mesonic operator
$X$ in the 4d effective action. In performing e.g. the $\mathbb{Z}_3$
quotient described above, there are three possible choices of
Chan-Paton phase for the E3-brane. This phase enters in the orbifold
projection on the E3-D3 fermion zero modes, and determines the
coupling of the survivors to one of the three bifundamentals $X_{12}$,
$X_{23}$, $X_{31}$ in the quotient theory. Therefore each of the three
possible E3-branes in the quotient theory lead to the insertion of one
of these baryonic operators. One can operate similarly to obtain
instantons with couplings to the other bifundamentals $Y_{i,i+1}$ or
$Z_{i,i+1}$ in the quotient theory.  Notice that single field
insertions for orbifold theories already provide the simplest
realization of the orbifolding procedure we are discussing in the
present section.

\medskip

Before continuing, let us make a few remarks on this simple example,
which generalize to arbitrary orbifolds. First notice that orbifolding
allows to deduce the appearance of baryonic operators from information
on the appearance of mesonic operators. Notice also that this examples
illustrates the above discussion on reducible vs recombinable cycles
in Section~\ref{generalities}. Consider a system of three E3-branes,
in the above $\mathbb{C}^3/\mathbb{Z}_3$ example, each with one of the
possible choices of Chan-Paton phase. The system of three E3-branes
can recombine into a single dynamical E3-brane which can move away
from the singularity. From our discussion in Section~\ref{mesonic},
such E3-brane leads to the insertion of the mesonic operator whose vev
parametrizes the E3-D3 distance. Indeed this agrees with our
discussion of reducible 4-cycles which can recombine. The three
different E3-branes lead to insertions of the operators $X_{12}$,
$X_{23}$, $X_{31}$. Taken together, the multi-instanton process they
generate leads to the insertion of their concatenation, namely the
mesonic operator $X_{12}X_{23}X_{31}$, which in fact corresponds to
the coordinate controlling the E3-D3 distance in the quotient theory.

Finally, the $\mathbb{Z}_3$ orbifold singularity also illustrates an
interesting feature in mapping the BPS operators under discussion, and
the corresponding 4-cycles, with the E3-brane instantons. Indeed, the
choice of Chan-Paton factor for a given 4-cycle can be described
geometrically as the choice of a holonomy at infinity for the
world-volume gauge field. Equivalently, the 3-cycle defining the base
of the conical 4-cycle is non-simply connected, and there is a
discrete choice of Wilson line. This implies that in the AdS theory,
for this 3-cycle there are different wrapped D3-brane states, which
correspond to different baryons. This is nicely correlated with the
existence, for such 4-cycle, of different E3-branes, coupling to
different bifundamentals. This provides another nice piece of
agreement between the E3-brane and the D3-brane viewpoint on BPS
operators of the field theory.

\medskip

Let us describe the extension of the above orbifolding procedure to
operators involving several fields. Consider for instance the
$\mathbb{C}^3$ theory with an E3-brane leading to the operator $X^2$,
which in fact corresponds to a system of two E3-branes recombinable
into a single one. Performing the quotient by the above $\mathbb{Z}_3$
action, one needs to specify the Chan-Paton action on the
E3-branes. Choosing the same Chan-Paton phase for both would lead to
operators of the form e.g. $X_{12}X_{12}$, for which the two fields
cannot be concatenated. This signals that the two E3-branes in the
quotient cannot be recombined (the recombination parameter has been
projected out by the quotient), hence it corresponds to an unavoidable
genuine 2-instanton process. On the other hand, choosing different
Chan-Paton phases leads to E3-branes generating insertions like
e.g. $X_{12}X_{23}$, namely long baryonic operators. These systems
correspond to E3-branes which admit a recombination into a single one,
and work as an overall single-instanton process. As already mentioned,
we focus on this kind of system, namely on E3-brane systems leading to
concatenated chains of bifundamentals.

This construction generalizes easily to obtaining the orbifold
descendants of general operators of the $\mathbb{C}^3$ theory. For
instance, operators like $XY$ lead to operators $X_{12}Y_{23}$,
$X_{23}Y_{31}$, $X_{31}Y_{12}$.  The choice of Chan-Paton phase on the
E3-brane system determines the endpoints of the chain of
bifundamentals (namely the baryonic charges of the operator). The
generalization should be clear. An important observation concerns the
absence of ordering ambiguities thanks to the use of F-term
relations. For instance, consider the operator $Y^2Z^2$ in
$\mathbb{C}^3$, and two of its possible descendants for a given choice
of Chan-Paton action e.g. $Y_{12}Y_{23}Z_{31}Z_{12}$ and
$Y_{12}Z_{23}Y_{31}Z_{12}$. These turn out to be identical upon using
the F-term equation for $X_{12}$, namely $Y_{23}Z_{31}=Z_{23}Y_{31}$,
as obtained from (\ref{supoz3}).

\subsection{Long baryonic couplings for general singularities from partial resolution}
\label{long-general}

In the previous section we have described the generation of long
baryonic operators for orbifold theories by E3-brane instantons. Since
partial resolutions of orbifold singularities can lead to non-orbifold
singularities, we may follow the effects of partial resolution on
E3-brane instantons in order to study long baryonic operators from
E3-brane instantons in non-orbifold singularities. In fact, since any
toric singularity can be regarded as the partial resolution of an
orbifold singularity (of sufficiently large order), partial resolution
can be used to obtain a general correspondence, for arbitrary toric
singularities, between single determinant BPS operators and E3-branes
on 4-cycles \footnote{In \cite{Imamura:2008fd} the couplings of
  flavour D7 branes were studied in the T-dual brane tiling picture,
  and a subset of the ``long'' couplings we discuss in this section
  were argued to exist.}. This correspondence is nicely correlated
with the map between BPS operators and 4-cycles defined by the AdS/CFT
correspondence.

\medskip

The main effects that a BPS operator (and the E3-brane instanton
generating it) can suffer in a process of partial resolution are the
following.
\begin{itemize}
\item All bifundamental fields in the chain defining the operator
  descend to fields in the resolved theory. The operator is unchanged
  and described by the same chain of fields in the resolved theory.

\item One of the bifundamental fields in the chain gets a vev.  The
  operator in the resolved theory is obtained by simply removing this
  bifundamental from the chain. Namely if the initial operator $({\cal
    {O}}_{\cal{P}})_{ab}=({\cal{O}}_{{\cal {P}}_1})_{ac} \Phi_{cd}
  ({\cal {O}}_{{\cal{P}}_2})_{db}$, with $\Phi_{dc}$ getting a vev,
  the operator in the resolved theory is $({\cal{O}'}_{\cal
    {P}'})_{ab}= ({\cal{O}}_{{\cal {P}}_1})_{ac} ({\cal
    {O}}_{{\cal{P}}_2})_{db}$. The operator remains ``single-determinant''
  since the vev for $\Phi_{cd}$ breaks of the two gauge factors $c,d$
  to the diagonal combination, so that the two sub-chains can be
  concatenated.

\item One of the bifundamental fields in the chain becomes massive by
  superpotential couplings and is not present in the resolved
  theory. Consider the operator $({\cal
    {O}}_{\cal{P}})_{ab}=({\cal{O}}_{{\cal {P}}_1})_{ac} \Phi_{cd}
  ({\cal {O}}_{{\cal{P}}_2})_{db}$, with $\Phi_{cd}$ becoming
  massive. To obtain the resolved theory the bifundamental is
  integrated out by using the F-term relations, which relate its value
  to some single determinant operator (possibly identically zero) say
  $\Phi_{cd}=({\cal{O}}_{{\cal{P}}_3})_{cd}$, involving fields that
  survive in the resolved theory. Since the BPS operators generated by
  the instantons should be understood modulo F-terms, the resulting
  operator in the resolved theory is simply $({\cal{O}'}_{\cal
    {P}'})_{ab}= ({\cal{O}}_{{\cal {P}}_1})_{ac}
  ({\cal{O}}_{{\cal{P}}_3})_{cd} ({\cal
    {O}}_{{\cal{P}}_2})_{db}$. This manifestly remains a single-determinant
  operator, i.e. a concatenated chain. In general, the replacement via
  F-term relation may require the replacement of a sub-chain in
  general longer than one bifundamental field.

\item The above two operations act quite trivially on the E3-brane,
  which still passes through the singularity after the process. There
  is however a situation where this geometrical property of the
  E3-brane changes. Notice that in a process of partial resolution
  some baryonic charges disappear. This implies that some baryonic
  operators lose their non-trivial charges and become mesonic in the
  resolved theory. For an operator $({\cal {O}}_{\cal{P}})_{ab}$ this
  happen when the groups $a,b$ are broken to the diagonal
  combination. The interpretation in terms of the E3-brane instanton
  is that the blowing-up process has grown a 2- or 4-cycle which
  separates the E3-branes from the D3-brane stack.

\end{itemize}

\medskip

\begin{figure}[ht!]
\begin{center}
  \includegraphics[scale=0.6]{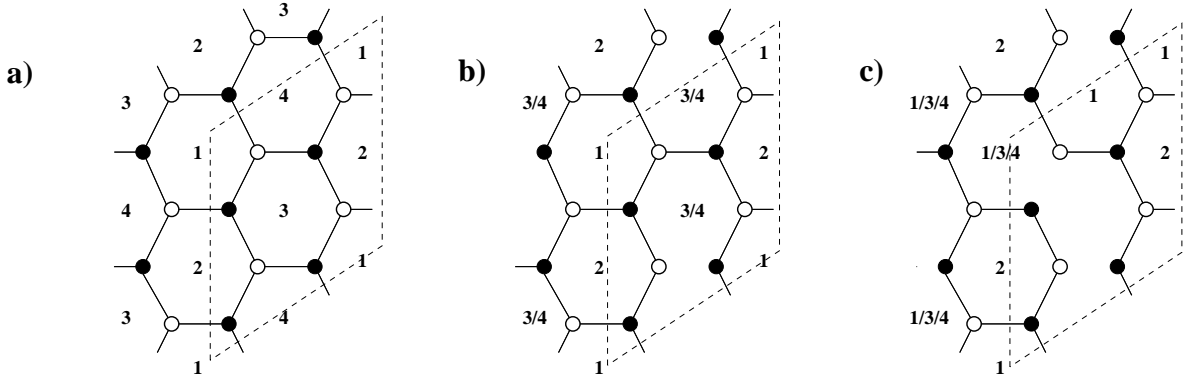}
  \caption{\small The dimer diagram for the
    $\mathbb{C}^3/(\mathbb{Z}_2\times \mathbb{Z}_2)$ theory (a) and
    its partial resolution to the SPP singularity (b) and the conifold
    (c). In order to keep the relation to the unresolved orbifold
    theory, we have not integrated out the bifundamental fields in
    di-valent nodes (mass terms in the superpotential).}
\label{dimerz2z2}
\end{center}
\end{figure}

Let us discuss these main features by considering an illustrative
example. Consider $\mathbb{C}^3/(\mathbb{Z}_2\times \mathbb{Z}_2)$,
with the orbifold generators associated to the twists $(\frac
12,-\frac 12,0)$ and $(0,\frac 12,-\frac 12)$. The gauge group of the
orbifold theory contains four factors, and the $\IC^3$ adjoints lead
to the bifundamental fields $X_{12}$, $X_{21}$, $X_{34}$, $X_{43}$,
$Y_{23}$, $Y_{32}$, $Y_{14}$, $Y_{41}$, and $Z_{31}$, $Z_{13}$,
$Z_{24}$, $Z_{42}$, in hopefully self-explanatory notation. The
superpotential has the structure $W\simeq XYZ-XZY$, with indices
distributed in all possible ways consistent with gauge invariance.
The dimer diagram is shown in Figure \ref{dimerz2z2}a. Figures b and c
provide the partial resolution to the SPP and the conifold, which we
are about to use, obtained by giving vevs to the fields $X_{34}$ for
the SPP, and to $X_{34}$, $Z_{31}$ for the conifold. The dimer is a
convenient tool to represent BPS operators, which correspond to paths
joining two faces (which are the same for mesonic operators), modulo
homotopy deformations (F-term relations). The effects described above
appear in this example as follows:
\begin{itemize}
\item The operator $X_{43}X_{34}Y_{41}$ in the orbifold theory
  descends to the operator $X_{43}Y_{41}$ in the SPP theory (which
  corresponds to a concatenated chain since the groups 3 and 4 become
  identified in the SPP theory).

\item Consider the operator $Z_{13}Y_{32}$ in the unresolved orbifold
  theory. The field $Z_{13}$ ends up as a massive one in the
  resolution to SPP, as is manifest in the dimer, where it enters
  a bi-valent node. It is however simple to deform the path in the SPP
  dimer to obtain the operator $X_{12}Z_{24}Y_{32}$ which is a
  concatenated chain (since 3 and 4 become identified) of fields
  massless in the SPP theory. This amounts to just using the F-term
  equivalence $Z_{13}X_{34}=X_{12}Z_{24}$ in the unresolved orbifold
  theory and replacing $X_{34}$ by its vev.

\item It is easy to find baryonic operators of the unresolved orbifold
  theory which become mesonic upon losing its baryonic charges in the
  partial resolution. The simplest example is just $X_{43}$, which is
  a mesonic operator in the SPP theory.
\end{itemize}

\medskip

It is easy to realize that the realization of general toric
singularities as partial resolution of orbifolds allows to reverse the
above line of argument. Namely one can show that any BPS operator
associated to a chain of bifundamentals in the non-orbifold theory can
be regarded as the resolved version of a chain of bifundamentals in
the orbifold theory. This construction produces the general map
between arbitrary single determinant BPS operators for toric field theories
and E3-branes instantons on 4-cycles producing them.

The correspondence can be easily argued to agree with the AdS/CFT map
between operators and wrapped D3-branes, given that the chain of
bifundamentals can be regarded as a monomial in the homogeneous
variables of the symplectic quotient of the construction, which
provide the defining equation for the 4-cycle on which to wrap the
E3-brane. This is precisely the map used in the AdS/CFT context.

Finally, let us point out an interesting crosscheck allowed by partial
resolution.  Considering the conifold theory in Figure
\ref{dimerz2z2}c, it is possible to resolve it completely to
$\mathbb{C}^3$ by giving a vev to any of the bi-fundamentals. This
partial resolution allows to recover long couplings in the
$\mathbb{C}^3$ theory by starting with long couplings of the conifold
theory. We have thus closed the circle and obtained a consistent
picture of all operators which can be generated using E3-brane
instantons.  Thus by orbifolding and partially resolving, one can
reach the general result that any single determinant BPS operator can be
generated from a suitable E3-brane instanton.

\subsection{The D6-brane mirror picture}
\label{mirror-picture}

\begin{figure}
\begin{center}
  \includegraphics[scale=0.6]{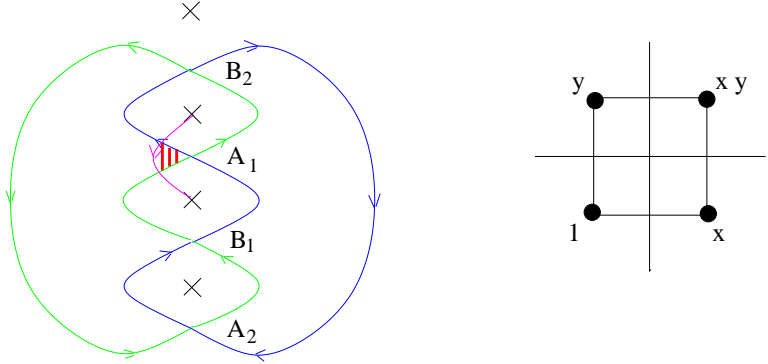}
  \caption{\small The mirror Riemann surface for the conifold, with
    punctures shown as crosses. The two 1-cycles in green and blue
    correspond to the two D3-brane gauge groups and the pink 1-cycle
    connecting the two punctures corresponds to the E3-brane
    instanton. The disk leading to the coupling $\alpha A_1 \beta $ is
    painted in red stripes. The instanton amplitude thus produces an
    insertion of the field $A_1$ (for the $N=1$ theory) in the
    D3-brane field theory. The right hand part of the figure shows the
    toric diagram for the conifold and its $(p,q)$-web.}
\label{shivercon}
\end{center}
\end{figure}

We can provide further arguments in favour of the couplings previously
discussed by using the mirror of the system of D3-branes at the
singularity. These are described in Appendix~\ref{mirror}, following
\cite{Feng:2005gw}. The mirror $\CW$ of a toric Calabi-Yau variety
$\mathcal{M}$ can be obtained starting from its toric diagram
\cite{Aharony:1997ju,Aharony:1997bh,Leung:1997tw}, as follows.  Assign
complex coordinates $x$, $y$ to the two axis of the toric diagram and
associate a monomial $x^iy^j$ to the point with coordinates $(i,j)$ in
the toric diagram. Define the polynomial $P(x,y)$ as the sum of all
these monomials with arbitrary complex coefficients \footnote{These
  complex coefficients parametrize the complex structure of the mirror
  manifold $\CW$, and they are mapped to the Kahler structure
  parameters of $\mathcal{M}$ under the mirror map. Their values are
  not relevant for our simplified discussion.}. The mirror variety
$\CW$ is defined by the equation $P(x,y)=uv$ where the coordinates
$u$, $v$ take values in $\mathbb{C}^*$. We can represent $\CW$ as a
double fibration over the complex plane with coordinate $z$
\begin{equation}
u v = z  \quad , \quad P(x,y) = z. 
\end{equation}
The first equation describes a $\mathbb{C}^*$ fibration, while the
second equation describes a fibration of a Riemann surface. The
structure of $\CW$ is essentially encoded in the latter fibration, and
in particular on the fiber over $z=0$.  The Riemann surface $P(x,y)=0$
has genus equal to the number of internal points of the toric diagram
and punctures corresponding to the external edges of the dual diagram
$\mathcal{C}^*$. For example in the conifold case we have a Riemann
surface that is topologically a sphere with four punctures, given by
the defining equation: $P(x,y)=1+x+y+xy$ (see Figure
\ref{shivercon}). In this mirror geometry, the gauge D3-branes
correspond to D6-branes wrapped on 3-cycles, which project on the
Riemann surface to non-trivial 1-cycles wrapping non-trivially around
the different punctures, in a way determined by the dimer diagram, see
Appendix~\ref{mirror}.  Intersections of these 1-cycles support
bifundamental chiral multiplets, while oriented disks defined by
different 1-cycles support worldsheet instantons leading to
superpotential couplings.

The mirror picture provides a nice geometric realization of the
euclidean instanton branes, their charged fermion zero modes, and
their couplings, as we now describe.  The mirror picture of the
E3-branes corresponds to E2-branes wrapped on non-compact special
Lagrangian 3-cycles, which project as 1-cycles in the Riemann surface,
escaping to infinity along two punctures. In fact, some of these
non-compact 3-cycles have appeared (describing the mirror of flavour
D7-branes) in \cite{Franco:2006es}. The intersection of the E3-brane
non-compact 1-cycle with the D3-brane compact 1-cycles lead to charged
fermion zero modes of the E3-brane instanton. Also, the disks bounded
by a given E3-brane non-compact 1-cycle and the D3-brane compact
1-cycles in the Riemann surface support worldsheet instantons
contributing to the couplings of the E3-brane instanton to a BPS
operators in the 4d field theory.

The explicit map between holomorphic 4-cycles and special Lagrangian
cycles is not know in general, thus in our analysis we consider a
shortcut. We start with a basic set of non-compact 1-cycles, which
correspond to E3-branes with fermion modes coupling to the basic
bifundamental chiral multiplets. In addition, we construct more
general E3-brane 1-cycles by combining basic 1-cycles which share a
common puncture. The physical interpretation is that one can form
bound states of the basic 1-cycles by giving vevs to fields in the
E3-E3' open string sector, triggering recombination of cycles. The
fermion zero modes and couplings of the resulting combined 1-cycle are
manifest from the Riemann surface picture, and agree with the naive
field theory analysis.  Let us explain this procedure using the
conifold example.

\medskip

Consider the 1-cycles corresponding to the E3-brane instantons
coupling to the elementary fields $A_i$, $B_j$ in the conifold
theory. As mentioned above, they are non-compact 1-cycles stretched
between punctures, and defining suitable disks involving the
corresponding bifundamental. The 1-cycles corresponding to E3-branes
with the desired structure of fermion zero modes and couplings, namely
$\alpha A_i \beta $, $\alpha B_j \beta $, are shown in Figure
\ref{shivercon}. Note that the pink 1-cycle on the Riemann surface
seems to define two disks, to its right and its left. However, only
the disk on the right has a well-defined boundary orientation, and can
really support a worldsheet instanton.

As discussed above, these basic cycles correspond to holomorphic
4-cycles (defined by the equations $x_r=0$ in the homogeneous
coordinates), and thus define supersymmetric 3-cycles in the mirror
picture. Consider for instance the two basic 1-cycles giving rise to
instantons coupling to $A_1$, $B_1$. In the type IIB picture, the two
instantons correspond to two 4-cycles, $x_1=0$ and $x_2=0$. This is a
situation where we argued that the two-instanton process can be
regarded as a limit of a one-instanton process, for a single E3-brane
instanton wrapping the recombined 4-cycle $x_1 x_2 +\epsilon= 0$, in
the limit $\epsilon \rightarrow 0$. Even in this limit, there is a
non-trivial contribution of the instanton, leading to the insertion of
the BPS operator $A_1B_1$. We can now show that this construction is
nicely reproduced using the mirror picture.
\begin{figure}
\begin{center}
  \includegraphics[scale=0.6]{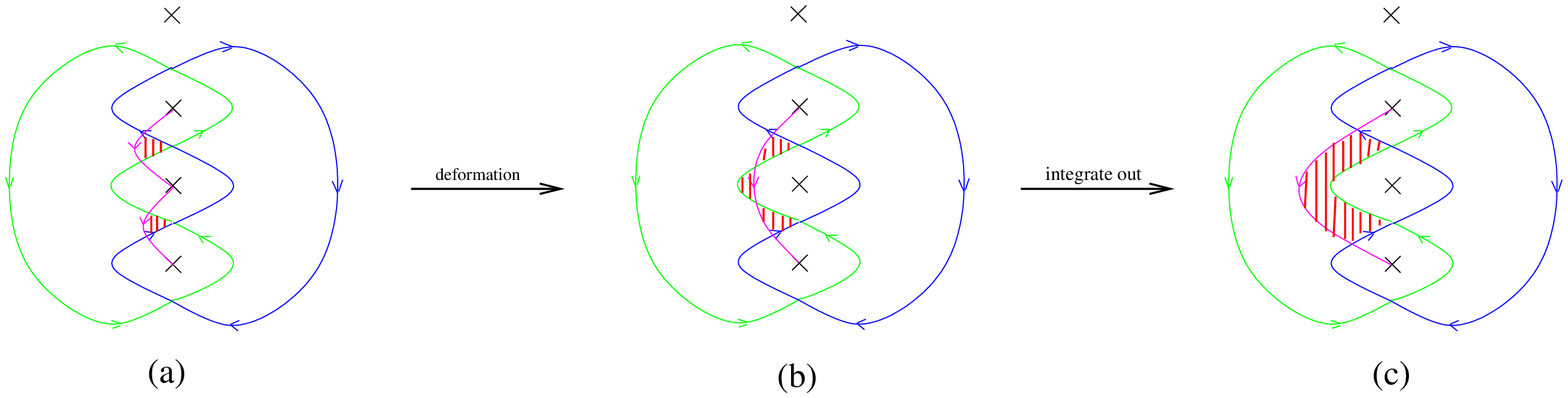}
  \caption{\small The two E2-brane instantons (a) can be recombined to
    a single E2-brane instanton (b) with coupling $\alpha A_1 \gamma +
    \gamma \delta + \delta B_1 \beta$. After integrating over the two
    modes $\gamma$, $\delta$ the coupling in (b) is equivalent to
    $\alpha A_1 B_1 \beta$, as shown pictorially in (c). Integration
    over the remaining modes leads to the appearance of the mesonic
    BPS operator $A_1B_1$ in the instanton non-perturbative
    amplitude.}
\label{shivercon2}
\end{center}
\end{figure}

Consider the two basic E3-brane 1-cycles in Figure~\ref{shivercon2}a,
describing instantons with fermion modes and couplings $\alpha A_1
\gamma$, $\delta B_1 \beta$.  The two 1-cycles share a common
puncture, corresponding to the fact that the IIB 4-cycles intersect
over a complex curve. This intersection supports an E3-E3' mode
$\phi_{33'}$ (for whose existence we choose appropriate boundary
conditions at infinity) with couplings $\gamma \phi_{33'}\delta$,
which follows pictorially from a disk in the Riemann surface. A vacuum
expectation value for this mode corresponds to the deformation
parameter $\epsilon$ mentioned above, leading to a single E3-brane
bound state, whose recombined 1-cycle is shown in
Figure~\ref{shivercon2}b. The triangle structure in the resulting
picture lead to couplings $\alpha A_1 \gamma + \gamma \delta + \delta
B_1 \beta$. One can thus integrate over the charged fermionic modes
$\gamma$, $\delta$ and obtain the coupling $\alpha A_1 B_1
\beta$. This corresponds pictorially to deforming the 1-cycle to
Figure \ref{shivercon2}c. Further integration over the remaining modes
leads to the insertion of $A_1B_1$ operators in the 4d instanton
amplitude.

In fact, even in the two-instanton process (with no recombination),
one can use the couplings in Figure ~\ref{shivercon2}a to saturate
over the zero modes $\gamma$, $\delta$, $\alpha$, $\beta$ and obtain
the insertion of the operator $A_1B_1$ from the two-instanton
process. In what follows, we will abuse language and use the above
pictorial representation to discuss processes involving
multi-instantons which can recombine into a single one, even when no
actual recombination is implied. The procedure can be describe using
two simple rules:
\begin{itemize}
\item{Two instantons coming in and out of the same puncture can be recombined into a single instanton.}
\item{One can deform 1-cycles to eliminate disks involving two intersections between the E3- and D3-brane 1-cycles (mass terms for non-chiral fermion modes). This correspond to integrating over the massive charged fermionic  modes.}
\end{itemize}

\begin{figure}[ht!]
\begin{center}
  \includegraphics[scale=0.6]{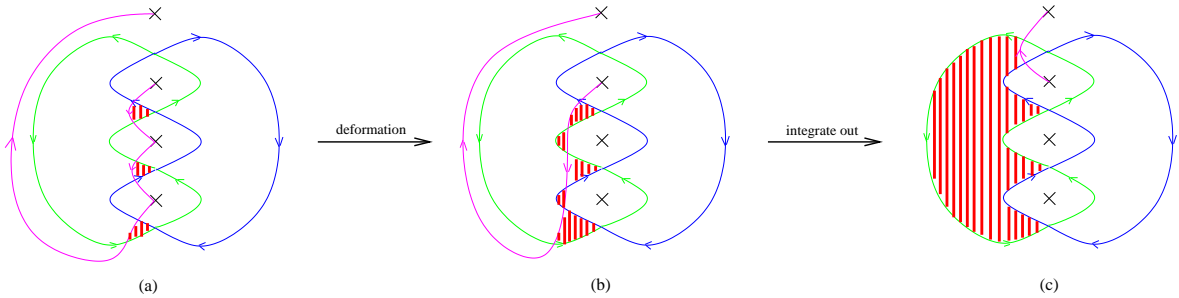}
\caption{\small  The three E2-brane instantons (a) can be recombined to a single E2-brane instanton (b) with coupling $\alpha A_1 \gamma + \gamma \delta + \delta B_1 \mu + \mu \nu + \nu A_2 \beta$. After integrating over the four modes $\gamma$, $\delta$, $\mu$, $\nu$ the coupling in (b) is equivalent to $\alpha A_1 B_1 A_2 \beta$, as shown pictorially in (c). Integration over the remaining modes leads to the appearance of the baryonic BPS operator  $ A_1 B_1 A_2 $ in the instanton non-perturbative amplitude.}
\label{shivercon3}
\end{center}
\end{figure}

Consider a further example, leading to an instanton coupling to the operator $A_1 B_1 A_2$. The pictorial representation, according to the above rules, is shown in  Figure~\ref{shivercon3}. The combined instanton system can be regarded as having the fermion modes and couplings $\alpha A_1 \gamma + \gamma \delta + \delta B_1 \mu + \mu \nu + \nu A_2 \beta$. Once we integrate over the four fermion modes $\gamma$, $\delta$, $\mu$, $\nu$ we obtain the equivalent coupling $\alpha A_1 B_1 A_2 \beta$ represented by the disk in the last figure. Integrating over these two charged zero modes give rise to the non perturbative insertion of the operator  $ A_1 B_1 A_2 $.

It is important to underline that there are other situations where the
multiple instantons behave as individual objects. These processes lead
to many additional zero modes, coming e.g. from the individual
goldstinos of the different instantons. Moreover, each instanton
carries its set of charged fermion zero modes, and integration over
them leads to the insertion of a BPS operator. Hence the
multi-instanton process leads to a ``multi-determinant'' BPS operator in the
field theory (and correspondingly, the boundary of the 4-cycles
corresponds to a multi-particle set of D3-branes). This also has a
nice interpretation in terms of the mirror geometry. E3-brane
instantons which cannot form a bound state are described by 1-cycles
which cannot recombine according to our above rules. Namely, they do
not share a puncture, or they do not have correct orientations when
they do. The structure of fermion modes and couplings from the Riemann
surface automatically leads to the insertions of ``multi-determinant'' BPS
operators.

\begin{figure}[ht!]
\begin{center}
  \includegraphics[scale=0.6]{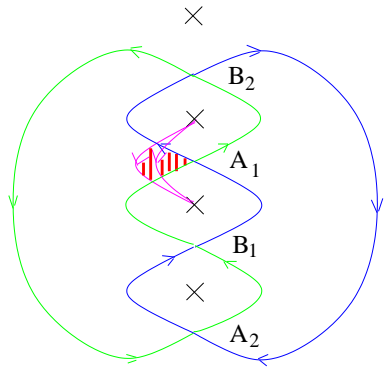}
\caption{\small The 1-cycles describing instantons with couplings $\alpha_1 A_1 \beta_1 + \alpha_2 A_1 \beta_2$. The two-instanton process leads to the insertion of the operator $A_1^2$ for $N=1$, or  $(\det ( A_1))^2$ for general $N$.}
\label{shivercon4}
\end{center}
\end{figure}

Let us consider a simple example. Consider the two 1-cycles in
Figure~\ref{shivercon4}. They describe two mutually BPS instantons,
each of them coupling to $A_1$, which cannot be recombined (due to
mismatch of orientations at the common punctures). In the IIB picture
this corresponds to the embedding equation $x_1^2=0$. Namely two
E3-brane instantons wrapped on the 4-cycle $x_1$. The system cannot
form a bound state, since the equation $x_1^2=0$ cannot be deformed
into a single one in a way consistent with the $\IC^*$ quotient. From
the mirror picture, we see that the instantons have fermion modes and
couplings $\alpha_1 A_1 \beta_1 + \alpha_2 A_1 \beta_2$. Integrating
over the charged zero modes $\alpha_i$, $\beta_j$ we have an insertion
of the operator $A_1^2$. Since the bi-fundamental structure of $A_1$
does not allow a concatenation of the two insertions, this corresponds
to a ``multi-determinant'' operator. Equivalently, considering the
theory for arbitrary $N$, the instantons generate the insertion of the
operator $(\det (A_1))^2$.

\section{BPS operators from E3-brane instantons:\\
  Extension to $N$ D3-branes}
\label{narb}

In the previous section we have shown a remarkable relation between
the E3-brane embeddings in $C(H)$ and the BPS operators in the quiver
theory of a single D3-brane at the singular point. Namely, once we
specify the geometric embedding and the holonomy at infinity of the
world volume gauge field, the E3 brane generates dynamically the
corresponding BPS operator. In this section we discuss the map between
E3-brane instantons and BPS operators for systems of $N$ D3-branes at
the toric singularity.

In passing from $N=1$ to general $N$, the spectrum of BPS operators
becomes much more complicated, and in general the correspondence
between BPS operators and string theory objects has to be studied at
the level of a generating set of BPS operators. For instance, as
explained in Section~\ref{genbpsop} a generating set is provided by
operators of the form (\ref{OPath}), namely the gauge invariant $N$
times symmetric products of concatenated chains of fields (analogous
to the ``single-determinant'' operators of the $N=1$ case). The set of all
BPS operators is obtained by taking products of these operators, and
linear combinations thereof. Let us focus on BPS operators given by
linear combination of operators of the form (\ref{OPath}), to which we
refer as `single-particle' for the moment. In the AdS/CFT setup, such
BPS operators correspond to D3-brane states in the Hilbert space of
the quantum mechanics in the space parametrized by the coefficients
$\{c_P\}$ in the defining equations of the holomorphic 4-cycles. A
particular basis of this Hilbert space is given by the states
(\ref{wavefunction}), dual to the operators (\ref{OPath}).

\medskip

In this section we show that E3-brane instantons provide, via the
computation of the non-perturbative field theory operators they
induce, a set of BPS operators which provide a generating set for all
`single particle' BPS operators. Namely any `single-particle' BPS
operator can be described as a linear combination of the basis
provided by the E3-brane instantons \footnote{Note that we do not
  imply that one can take linear combinations of E3-brane instantons
  to achieve an arbitrary BPS operator in the field theory.}. At the
level of the AdS/CFT setup, the D3-brane states corresponding to the
E3-brane instantons are those dual to determinant operators, as
discussed in Section~\ref{determinants}, and provide a basis of the
same Hilbert space spanned by the states (\ref{wavefunction}), as
shown in Section~\ref{veronese}. Therefore, although there is no
one-to-one correspondence between BPS operators and E3-brane
instantons, the E3-brane instantons do provide a generating set of BPS
operators. This is enough to support the view that the correspondence
between E3-brane instantons and BPS operators underlies the familiar
one-to-one map between quantum D3-brane states and BPS operators.

\subsection{The determinant operators}
\label{determinants}

The considerations in Section~\ref{single} for the $N=1$ case allow a
simple generalization to arbitrary $N$. Using the couplings between
E3-brane instantons and D3-branes in the $N=1$ case, we may increase
the range of D3-brane Chan-Paton indices to obtain the worldvolume
fermion modes and couplings of the general $N$ system. Namely, for the
operator ${\cal {O}}_{\cal{P}}$ corresponding to any concatenated
chain of bifundamentals described by a path ${\cal {P}}$ in the dimer,
there is an E3-instanton wrapped on a holomorphic 4-cycle with charged
fermion modes and couplings $\alpha {\cal{O}}_{\cal {P}} \beta$. Here
the modes $\alpha$ and $\beta$ transform in the $\antifund$ and
$\fund$, of the gauge groups where the path ${\cal {P}}$ start and
end, respectively (and which are the same for a mesonic
operator). Integration over these fermion modes leads to insertion of
the field theory BPS operator \beqa
\label{nnperi}
\det \,{\cal{O}}_{\cal{P}}\, =\,  \epsilon_{\tilde{p}_1,\ldots,\tilde{p}_N} \, \epsilon_{k_1,\ldots,k_N} \, ({\cal {O}}_{{\cal{P}}})_{p_1\tilde{k}_1}... ({\cal {O}}_{{\cal {P}}})_{p_N\tilde{k}_N}\, . 
\eeqa 
Equivalently, for each possible 4-cycle, or equivalently for each choice of monomials $\{c_P\}$ in the defining equation, there is a BPS operator $\det{\cal {O}}_{\cal{P}}$.

The fact that this mechanism only generates determinant operators
might suggest that such operators cannot generate the whole set of BPS
operators, in particular operators of the form (\ref{OPath}) with
different entries ${\cal {O}}_{\cal {P}_I}$. However, the fact that we
have an operator for each possible choice of 4-cycle (out of an
infinite set, parametrized by the $c_P$) implies that the set of
determinant operators generates the complete Hilbert space of baryonic
operators, as we show in the next section. Note that the generating
set of operators provided by these D3-brane states associated with E3
branes is unfamiliar from the CFT viewpoint, since it involves linear
superpositions of operators with different conformal dimensions.

\subsection{The space of general BPS operators and the Veronese map}
\label{veronese}

Let us introduce the shorthand notation
\begin{equation}
({\cal{O}}_{{\cal{P}}_1},...,{\cal{O}}_{{\cal{P}}_N})\, =\,
\epsilon_{\tilde{p}_1,\ldots,\tilde{p}_N} \, \epsilon_{k_1,\ldots,k_N}
\, ({\cal {O}}_{{\cal{P}}_1})_{p_1\tilde{k}_1}... ({\cal {O}}_{{\cal
    {P}}_N})_{p_N\tilde{k}_N}
\end{equation}
The set of operators for all possible choices of paths $\{ {\cal
  {P}}_I\}$ forms a basis of BPS operators. Let us consider the
question of whether it is possible to reproduce the above operators by
considering linear combinations of the determinant operators
(\ref{nnperi}).

\medskip

Consider first the simple example of the conifold with just two
colours $N=2$, and the reduced problem of constructing all the
possible operators of baryonic charge 2 involving just by the two
chains of bifundamentals $A$, $ABA$. One basis for these operators is
given by
\begin{equation}
e_1=(A,A) \hbox{ , } e_2=(ABA,ABA) \hbox{ , } e_3=(A,ABA)
\end{equation}
and the general operator corresponds to a linear combination
thereof. We denote the set of these by ${\cal {M}}_2$.

The operators $A$, $ABA$ correspond to two specific monomials of the
homogeneous coordinates of the conifold, that for simplicity we just
call $x$, $y$: $A \rightarrow x$, $ABA \rightarrow y$. These monomials
define sections of a non-trivial line bundle over the conifold. The
generic section spanned by them is $f=a x+by$, where $a,b$ are complex
coefficients, in the defining equation of the 4-cycle $f=0$, dual to
the corresponding operator. Consider the determinant operator
generated by an E3-brane instanton wrapped on this 4-cycle, namely
\begin{eqnarray}\label{opab}
  O(a,b)&=&\det(f)\,=\, (f,f)\, =\, a^2\, (x,x)\, +\, 2ab\,(x,y)\, +\, b^2\,(y,y)\, = \nonumber \\
  & & a^2\, (A,A)\, +\, 2ab\, (A,ABA)\, +\, b^2\, (ABA,ABA)
\end{eqnarray}
Hence E3-brane instantons lead to operators $O(a,b)$ for arbitrary
choices of $a,b$. In order to show that this set is generating, we
need to show that one can choose particular values of $(a,b)$ to
obtain three linearly independent operators generating
${\cal{M}}_2$. In this case it is easy to find that e.g.
\begin{equation}
l_1=O(1,0)=e_1 \hbox{ , } l_2=O(0,1)=e_2 \hbox{ , } l_3=O(1,1)=e_1+e_2+2 e_3
\end{equation}
provide a basis of the same space of operators ${\cal {M}}_2$. 

\medskip

In order to generalize the above construction to arbitrary $N$, it is
convenient to express it in more geometric terms. Since the equation
$f=a x+ b y=0$ is invariant under complex rescalings of $a,b$, the set
of such equations is a $\mathbb{P}^1$, with homogeneous coordinates
$a,b$. Similarly, the BPS operators are given by linear combinations
of the $e_i$, $\sum z_i e_i$, up to overall rescaling of the $z_i$,
namely they are parametrized by a $\mathbb{P}^2$. The computation of
the BPS operator corresponding to an E3-brane instanton on a 4-cycle
defined by $(a,b)$ defines a map
\begin{eqnarray}
v_1 &:& \mathbb{P}^1 \rightarrow \mathbb{P}^2 \nonumber\\
    & & [a;b] \rightarrow [a^2;ab;b^2]
\end{eqnarray}
This is an example of a well know construction in algebraic geometry
called the (degree~2) Veronese embedding. The image set in
$\mathbb{P}^2$ is given by the degree 2 curve
\begin{equation}
  \label{veronese-two}
  z_1 z_3- z_2^2=0.
\end{equation}

The set of operators $O(a,b)$ will form a basis of ${\cal {M}}_2$ if
there exist at least three points in $\mathbb{P}^1$ such that the
vectors $l(a,b)=a^2 e_1 + ab e_2 + b^2 e_3$ form a basis of
$\mathbb{C}^3$. In geometric terms, a basis will not be obtained only
if the image $v_1(\mathbb{P}^1)$ is contained in a hyperplane in
$\mathbb{P}^2$. It is a familiar result of algebraic geometry that the
Veronese curve is indeed not contained in any hyperplane \footnote{In
  this simple case the statement can be easily seen to be true, since
  it amounts to the trivial fact that it is not possible to rewrite
  the quadratic equation (\ref{veronese-two}) into a linear equation
  of the form $\sum c_i z_i = 0$, for any constant $c_i$.}. Hence the
set of operators $O(a,b)$ forms a generating set.

\medskip

Let us pass on to the general case. Using the tools and the intuition
we have just developed we can show that the set of BPS operators
induced by all possible E3-brane instantons form a generating set of
all BPS operators of the quiver gauge theory for general $N$
\footnote{As explained in the previous section, we need to consider
  processes involving multiple instantons, and they correspond to
  multi-particle D3-brane states in AdS/CFT. For simplicity we
  restrict to single-particle BPS operators, since they can generate
  the complete set of all BPS operators.}.

We start the discussion explaining the general form of the Veronese
map, which plays a prominent role in the argument, and which is a
simple generalization of the $N=2$ discussion above. The general
Veronese map is an embedding of $\mathbb{P}^m$ in $\mathbb{P}^n$
defined as follows. Consider $\mathbb{P}^m$ parametrized by
homogeneous coordinates $[u_0,...,u_m]$. The set of degree $N$
homogeneous polynomials in these coordinates
\begin{equation}
\sum_{i_0+\ldots + i_m=N} \, w_{i_0...i_m}\, u_0^{i_0}\ldots u_m^{i_m}
\end{equation}
defines a vector space of dimension $\binom{m+N}{N}$, with coordinates
$w_{i_0...i_m}$. Let us take as the target of our Veronese map
$\mathbb{P}^n$, with $n=\binom{m+N}{N}-1$. This parametrizes, as
above, the set of homogeneous polynomials modulo an overall
rescaling. The degree $N$ Veronese map is obtained by considering the
$N^{th}$ power of a general monomial in the $u_k$, namely it is
defined by the map
\begin{eqnarray}
v_N &:& \mathbb{P}^m \rightarrow \mathbb{P}^n \nonumber\\
    & &  w_{i_0...i_m} = u_0^{i_0}...u_m^{i_m}
\end{eqnarray} 
for $i_0+...+i_m = N$. The resulting Veronese variety
$v_N(\mathbb{P}^m) \in \mathbb{P}^n$ can also be described by the
following set of quadrics, which follow from the specific form of the
embedding:
\begin{equation}
w_{i_0...i_m}w_{j_0...j_m}=w_{k_0...k_m}w_{l_0...l_m}
\end{equation}
whenever $i_0+j_0=k_0+l_0,...,i_m+j_m=k_m+l_m$. It is a general result
that the variety $v_N(\mathbb{P}^m) \in \mathbb{P}^n$ is not contained
in any linear subspace of $\mathbb{P}^n$.

\medskip

The application of this result to our problem of mapping of BPS
operators for the conifold case should be clear by now. In fact, it
can be used to solve the mapping problem for arbitrary toric
singularities, as we now argue. A generic toric variety can be
described as a symplectic quotient of $\mathbb{C}^d$ by the action of
an abelian group $K$ ($K \sim (\mathbb{C}^*)^{d-3}\times \Delta$ where
$\Delta$ is some abelian discrete group). Denoting $x_1,...,x_d$ the
homogeneous coordinates, the supersymmetric 4-cycles on which one can
wrap E3-brane instantons are given by equations
\begin{equation}
f\, =\, \sum_{i_1, \ldots,i_d} c_{i_1...1_d} x_1^{i_1}...x_d^{i_d}\, =\, 0
\end{equation}
that transform homogeneously under $K$. Let us momentarily restrict
the infinite set of coefficients $a_P=c_{i_1...i_d}$ to a finite set
of $m+1$. Then the set of holomorphic 4-cycles parametrizes a
$\mathbb{P}^m$ with homogeneous coordinates
$[\,\{a_P\}]=[a_1,\ldots,a_{m+1}]$. As discussed in Section
\ref{single}, every monomial $x_1^{i_1}...x_d^{i_d}$ is associated to
a concatenated chain of bifundamentals in the quiver field theory
\footnote{Here we are simplifying slightly, and restricting ourselves
  to the single particle case.}, defining an operator ${\cal
  {O}}_{\cal{P}}$. Increasing the range of Chan-Paton indices to
general $N$, an E3-brane wrapped on the holomorphic 4-cycle leads to
fermion zero modes and couplings $\alpha {\cal {O}}_{\cal{P}} \beta
$. Integration over fermion zero modes leads to the BPS operator $\det
{\cal {O}}_{\cal{P}} = ({\cal {O}}_{\cal{P}},\ldots,{\cal
  {O}}_{\cal{P}})$ in the 4d field theory. Expanding this determinant,
i.e. taking all possible degree $N$ products of the monomials
contained in $f$ (or rather its field theory translation), we obtain a
linear combination of the set of operators of the form
(\ref{OPath}). In this way the set of operators obtained by all
possible embeddings of the instanton is described by the degree $N$
Veronese embedding from $\mathbb{P}^m$ to $\mathbb{P}^n$, and we have
argued above that such a embedding spans a base of all possible
operators.

In order to complete the argument we just need to remove the cutoff
$m$, a step which does not modify the conclusions.

\section{Conclusions and Outlook}\label{conclu}

In this paper we have discussed the field theory operators on the
worldvolume theory of systems of D3-branes at toric singularities
induced by E3-brane instantons wrapped on holomorphic 4-cycles on the
Calabi-Yau geometry. We have argued that the resulting correspondence
between E3-branes on 4-cycles and BPS baryonic operators in the quiver
theory underlies and explains the AdS/CFT correspondence between
wrapped D3-brane states on AdS and BPS operators on the boundary
theory. Let us suggest some further applications and possible future
research directions.

\medskip

We have described the correspondence between E3-brane instantons and BPS 
operators in terms of a generating set of the latter. Namely any BPS 
operator can be written as a combination of the BPS operators directly 
induced by E3-brane instantons.
This operation has a well-defined meaning in the AdS/CFT context, where 
the wrapped D3-branes from the E3-brane instantons form a complete set of 
quantum states of the Hilbert space dual to the set of BPS operators. 
Since the operation of taking linear combination has a physical meaning 
for the wrapped D3-brane states, there is a one-to-one map between wrapped 
D3-branes and BPS operators. 

It would be interesting to explore physical realizations of this map at 
the level of the E3-brane instantons. One tantalizing possibility, 
suggested by the structure of the operators (\ref{OPath}) and its dual 
states (\ref{wavefunction}), is considering fractional instantons.
In gauge theories, fractional instantons are physical objects whose action 
and number of fermion zero modes is a (typically $1/N$) fraction of those 
for a genuine instanton. They have been suggested (see e.g. 
\cite{Brodie:1998bv}) as responsible for the gaugino condensate of $SU(N)$ 
super-Yang-Mills (or more generally for the non-perturbative 
superpotential of SQCD for $N_f<N_c-1$). They have also been proposed to 
play a prominent role in the strong coupling dynamics of more general 
supersymmetric gauge theories. Although the physical interpretation of 
fractional non-gauge D-brane instantons it far from clear, it is tempting 
to propose that a genuine E3-brane instanton is made up of $N$ fractional 
instantons, each coupling to a particular concatenated chain of 
bifundamentals along a dimer path ${\cal{P}}$. In such interpretation, the 
BPS operator (\ref{OPath}) would correspond to a set of $N$ fractional 
E3-brane instantons, each coupling to a different path ${\cal {P}}_I$, 
$I=1,\ldots, N$.  We leave this as an open direction for further research.

\medskip

A second interesting tool to attempt the formalization of a one-to-one map 
between BPS operators and E3-branes is provided by the master space of the 
supersymmetric quiver theory introduced in 
\cite{Forcella:2008bb,Forcella:2008eh}. It describes the 
set of gauge invariant BPS operators of the D3-brane field theory, and has 
a systematic algebraic geometry description for the case of field theories 
on D3-branes at singularities. Moreover, it helps in reducing the problem 
of counting operators for general $N$ to a plethystic exponentiation of 
the counting problem for the $N=1$ theory. Since for the $N=1$ theory 
there exists a nice explicit map between BPS operators and E3-brane 
instantons, it sounds plausible that the master space can provide a 
meaningful physical one-to-one map between BPS operators and E3-branes for 
general $N$. 

\medskip

Finally we would like to emphasize that our discussions of the field
theory operators induced by general E3-brane instantons should have
interesting applications to model building. Indeed, in the
construction of realistic particle models on D3-branes at
singularities it is natural to look for potential sources for
particular interesting field theory operators, which are forbidden in
perturbation theory and could be generated by instantons. Our tools
can be used to propose precisely the kind of E3-brane instanton
required for a given field theory operator. Thus our work allows for a
broad generalization of the work in \cite{Ibanez:2007tu}. Another
recent model building proposal related to our work is D-brane
instanton mediated supersymmetry breaking in \cite{Buican:2008qe}.

We expect much progress on the systematic understanding of E3-brane 
instantons and their corresponding BPS operators for these and other new 
directions.

\section*{Acknowledgements}
It is a pleasure to thank F. Ferrari, S. Franco, A. Hanany,
L. E. Ib\'a\~nez, D. Rodr\'iguez-G\'omez and A. Zaffaroni for
enlightening discussions. I.G.-E. thanks the CERN Theory division for
hospitality, and N. Hasegawa for kind support. This work has been
supported by the European Commission under RTN European Programs
MRTN-CT-2004-503369, MRTN-CT-2004-005105, by the CICYT (Spain) and the
Comunidad de Madrid under project HEPHACOS P-ESP-00346. The work of
I.G.-E. was financed by the Gobierno Vasco PhD fellowship program.
D.~F.~is supported in part by INFN and the Marie Curie fellowship
under the programme EUROTHEPHY-2007-1.

\appendix

\section{Short review of dimer models}
\label{dimerreview}

In this appendix we would like to present a short review of the main
aspects of dimer model technology that enter in our analysis
above. For a more in-depth discussion, the reader is advised to
consult the excellent reviews \cite{Kennaway:2007tq,Yamazaki:2008bt}.

\subsection{Quiver gauge theories and dimer diagrams}
\label{quiver}

\begin{figure}[!hbp]
\centering
\psfrag{w}{$W = - X_{21}X_{12}X_{23}X_{32} + X_{32}X_{23}X_{34}X_{43}$}
\psfrag{x}{$- X_{43}X_{34}X_{41}X_{14} + X_{14}X_{41}X_{12}X_{21}$}
\psfrag{QUIVER}{Quiver}
\psfrag{DIMER}{Dimer}
\includegraphics[scale=0.70]{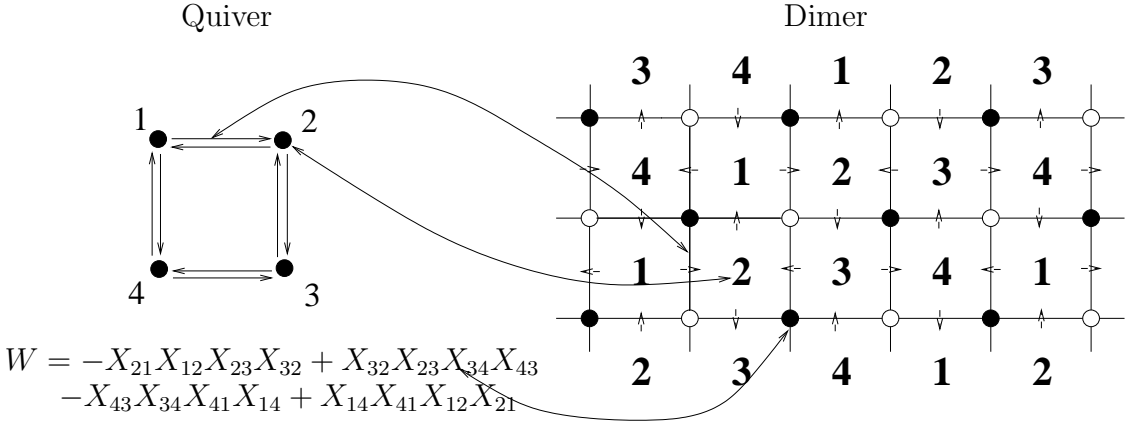}
\caption{\small Quiver and dimer for a $\mathbb{Z}_2$ orbifold of the conifold. 
Faces in the 
dimer correspond to gauge groups, edges correspond to bifundamentals and 
each vertex corresponds to a superpotential term. Edges have an 
orientation determined by the coloring of the adjacent nodes.}
\label{dconidimer_quiver}
\end{figure}

The gauge theory of D3-branes probing toric threefold singularities is
determined by a set of unitary gauge factors (of equal rank in the
absence of fractional branes, which we do not consider for the
moment), chiral multiplets in bi-fundamental representations, and a
superpotential given by a sum of traces of products of such
bi-fundamental fields.  The gauge group and matter content of such
gauge theories can be encoded in a quiver diagram, such as that shown
in Figure~\ref{dconidimer_quiver}a, with nodes corresponding to gauge
factors, and arrows to bi-fundamentals.  The superpotential terms
correspond to closed loops of arrows, but the quiver does not fully
encode the superpotential.

Recently it has been shown that all the gauge theory information,
including the gauge group, the matter content and the superpotential,
can be encoded in a so-called brane tiling or dimer graph
\cite{Hanany:2005ve,Franco:2005rj} \footnote{The brane tiling / dimer
  diagram can be dualized to an improved quiver diagram, the periodic
  quiver, which also encodes all this information.}. This is a tiling
of $\mathbb{T}^2$ defined by a bi-partite graph, namely one whose
nodes can be colored black and white, with no edges connecting nodes
of the same color. The dictionary associates faces in the dimer
diagram to gauge factors in the field theory, edges with
bi-fundamental fields (fields in the adjoint in the case that the same
face is at both sides of the edge), and nodes with superpotential
terms. The bi-partite character of the diagram is important in that it
defines an orientation for edges (e.g. from black to white nodes),
which determines the chirality of the bi-fundamental fields. Also, the
color of a node determines the sign of the corresponding
superpotential term.

The explicit mapping between this bipartite graph and the gauge theory, 
is illustrated in one example in Figure~\ref{dconidimer_quiver}.
%

\subsection{Dimer diagrams and the mirror Riemann surface}
\label{mirror}

\begin{figure}[!htp]
\begin{center}
\psfrag{Sigma}{$\Sigma$}
\includegraphics[scale=0.5]{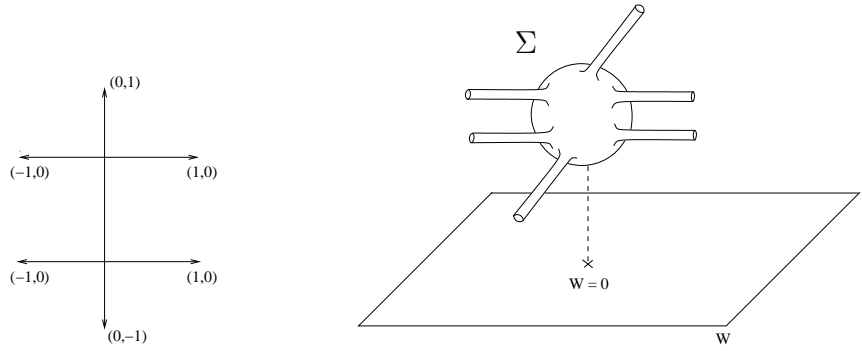}
\caption{\small a) An example of a web diagram (for the complex cone over 
$\mathbb{F}_0$); b) the corresponding Riemann surface $\Sigma$ in the mirror 
geometry.}
\label{thickweb}
\end{center}
\end{figure}

In \cite{Feng:2005gw} it was shown how, using mirror symmetry, it is
possible to relate in a useful and explicit way the gauge theory on
the D3 branes on the singularity and dimer models. Let us summarize
the results of that paper here. The mirror geometry to a Calabi-Yau
singularity $\mathcal{M}$ is specified by a double fibration over the
complex plane W given by
\begin{eqnarray}
 W \ &=&\ P(z,w)\\
 W \ &=& \ uv
\end{eqnarray}
with $w,z \ \in \mathbb{C}^*$ and $u,v \ \in \mathbb{C}$.\
Here $P(z,w)$ is the Newton polynomial of the toric diagram of  
$\mathcal{M}$. The surface $W = P(z,w)$ describes a genus $g$ 
Riemann surface $\Sigma_W$ with punctures, fibered over $W$. The genus
$g$ equals the number of internal points of the toric diagram. The fiber 
over $W=0$, denoted simply $\Sigma$, will be important for our purposes. 
It corresponds to a smooth Riemann surface which can be thought of as a 
thickening of the web diagram 
\cite{Aharony:1997ju,Aharony:1997bh,Leung:1997tw} dual to the toric 
diagram, see Figure \ref{thickweb}.
%

\begin{figure}[!htp]
\begin{center}
\epsfxsize=8cm
\hspace*{0in}\vspace*{.2in}
\epsffile{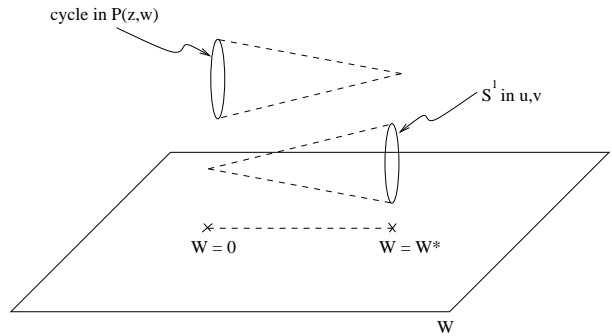}
\caption{\small Structure of the non-trivial 3-cycles in the geometry 
$\mathcal{W}$. They are constructed by fibering over the segment 
joining $W=0$ and $W=W^*$, the $S^1$ in the $uv$ fiber (degenerating at 
$W=0$) times the 1-cycle in the $P(z,w)$ fiber degenerating at $W=W^*$.}
\label{s3fiber2}
\end{center}
\end{figure}

At critical points $W=W^*$, a cycle in $\Sigma_W$ degenerates and pinches 
off. Also, at $W=0$ the $S^1$ in $W=uv$ degenerates. One can use these 
degenerations to construct non-trivial 3-cycles in the mirror geometry as 
follows. Consider the segment in the $W$-plane which joins $W=0$ with one 
of the critical points $W=W^*$, and fiber over it the ${\bf S}^1$ in $W=uv$ 
times the 1-cycle in $\Sigma_W$ degenerating at $W=W^*$,
see Figure \ref{s3fiber2}. The result is a 3-cycle with an ${\bf S}^3$ 
topology. The number of such degenerations of $\Sigma_W$, and hence the 
number of such 3-cycles, is given by twice the area of the toric diagram.

Mirror symmetry specifies that the different gauge factors on the 
D3-branes in the original singularity arise from D6-branes 
wrapping the different 3-cycles. The 3-cycles on which the D6-branes wrap 
intersect over $W=0$, precisely at the intersection points of the 
1-cycles in $\Sigma_{W=0}$. Open strings 
at such intersections lead to the chiral bi-fundamental fields. Moreover, 
disks in $\Sigma$ bounded by pieces of different 1-cycles lead to 
superpotential terms generated by world-sheet instantons.

Hence, the structure of the 3-cycles, and hence of the gauge theory,
is determined by the 1-cycles in the fiber $\Sigma$ over $W=0$. This
structure, which is naturally embedded in a $\mathbb{T}^3$ (from the
$\mathbb{T}^3$ fibration structure of the mirror geometry), admits a
natural projection to a $\mathbb{T}^2$, upon which the 1-cycles end up
providing a tiling of $\mathbb{T}^2$ by a bi-partite graph, which is
precisely the dimer diagram of the gauge theory.

\begin{figure}[!htp]
\begin{center}
\epsfxsize=10cm
\hspace*{0in}\vspace*{.2in}
\epsffile{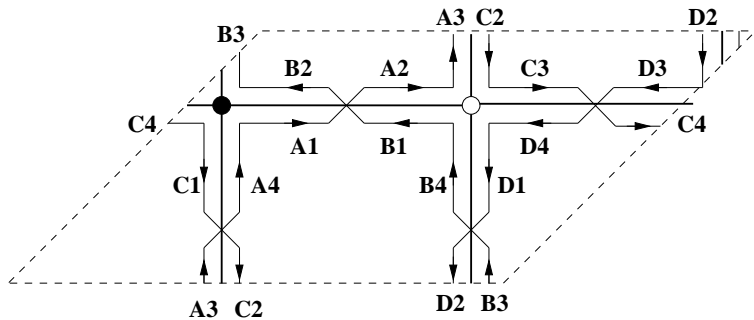}
\caption{\small Dimer of the conifold with the corresponding zig-zag paths.}
\label{conidimer_zigzag}
\end{center}
\end{figure}

This last process is perhaps better understood (and of more practical use) 
by recovering the Riemann surface $\Sigma$ from the dimer diagram of the 
gauge theory, as follows. Given a dimer diagram, one can define 
zig-zag paths (these, along with the related rhombi paths, were
introduced
in the mathematical literature on dimers in \cite{kenyon1,kenyon2},
and applied to the quiver gauge theory context in
\cite{Hanany:2005ss}), as paths composed of edges, and which turn
maximally to the right at e.g. black nodes and maximally to the left at 
white nodes. They can be conveniently shown as oriented lines that cross 
once at each edge and turn at each vertex, as shown in Figure 
\ref{conidimer_zigzag}.

Notice that at each edge two zig-zag paths must have opposite
orientations. For dimer models describing toric gauge theories, these
zig-zag paths never intersect themselves and they form closed
loops wrapping $(p,q)$ cycles on the $\mathbb{T}^2$. This is shown for
the conifold in Figure \ref{conidimer_zigzag} where the zig-zag paths
A, B, C and D have charges (0,1), (-1,1), (1,-1), (0,-1) respectively.

\begin{figure}[!htp]
\begin{center}
\epsfxsize=10cm
\hspace*{0in}\vspace*{.2in}
\epsffile{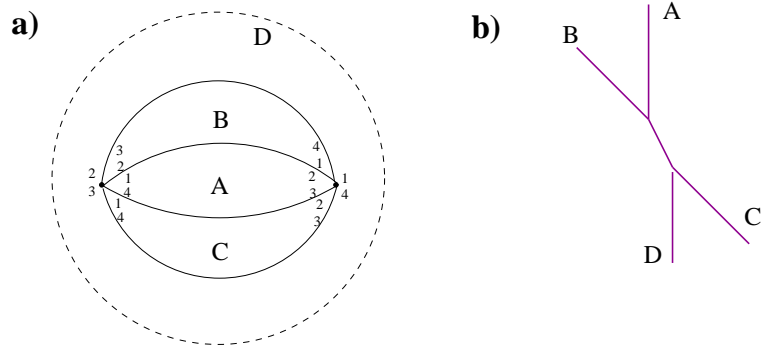}
\caption{\small a) Tiling of the Riemann surface (which is 
topologically a sphere, shown as the complex plane) for the case of 
D3-branes at a conifold singularity. b) The web diagram, providing a 
skeleton of the Riemann surface, with asymptotic legs 
corresponding to punctures (and hence to faces of the tiling of $\Sigma$, 
and zig-zag paths of the original dimer diagram).}
\label{coniriem}
\end{center}
\end{figure}

As shown in \cite{Feng:2005gw}, the zig-zag paths of the dimer diagram 
associated to D3-branes at a singularity lead to a tiling of the Riemann 
surface $\Sigma$ in the mirror geometry. Specifically, each zig-zag path 
encloses a face of the tiling of $\Sigma$ which includes a puncture, and 
the $(p,q)$ charge of the associated leg in the web diagram is the $(p,q)$ 
homology charge of the zig-zag path in the $\mathbb{T}^2$. The touching of 
two of these faces in the tiling of $\Sigma$ corresponds 
to the coincidence of the corresponding zig-zag paths along an edge of the 
dimer diagram. The tiling of $\Sigma$ for the conifold is shown in Figure 
\ref{coniriem}a, while the corresponding web diagram is shown in Figure 
\ref{coniriem}b.

\begin{figure}[!htp]
\begin{center}
\epsfxsize=10cm
\hspace*{0in}\vspace*{.2in}
\epsffile{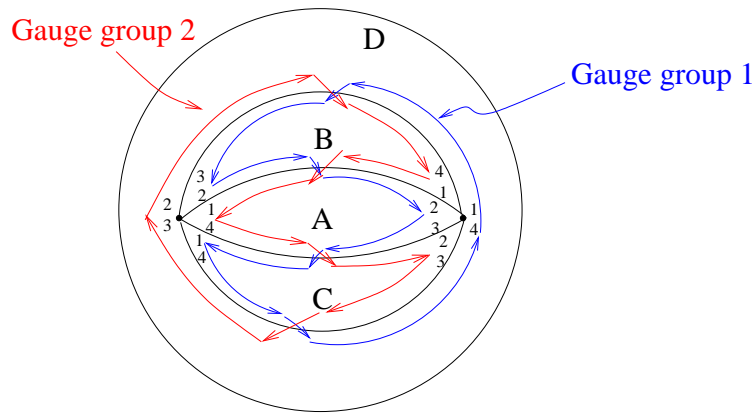}
\caption{\small Tiling of the Riemann surface for the case of D3-branes 
at a conifold singularity, with the 1-cycles corresponding to the two 
gauge factors (shown as zig-zag paths of the tiling of $\Sigma$).}
\label{conisurf}
\end{center}
\end{figure}

The dimer diagram moreover encodes the 1-cycles in the mirror Riemann
surface, associated to the different gauge factors in the gauge
theory.  Consider a gauge factor associated to a face in the dimer
diagram. One can consider the ordered sequence of zig-zag path pieces
that appear on the interior side of the edges enclosing this face. By
following these pieces in the tiling of $\Sigma$ one obtains a
non-trivial 1-cycle in $\Sigma$ which corresponds precisely to that
used to define the 3-cycle wrapped by the mirror D6-branes carrying
that gauge factor. Using this map, it is possible to verify all dimer
diagram rules (edges are bi-fundamentals, nodes are superpotential
terms) mentioned at the beginning. The non-trivial 1-cycles in the
mirror Riemann surface for the case of the conifold are shown in
Figure \ref{conisurf}.

\subsection{Resolution of the singularity}

In the discussion in the main text, we are particularly interested in
seeing how does resolution of the singularity appear on the dimer
model description. This resolution is expected to be represented in
terms of the gauge theory as a Higgsing of some fields, in such a way
that the low energy theory after Higgsing is the gauge theory of
branes placed at the two daughter singularities. As shown in
\cite{GarciaEtxebarria:2006aq}, it is possible to give a simple and
beautiful recipe for understanding which fields get vevs using the
mirror description of the system. Let us summarize the main points of
the procedure here.

\medskip

In terms of the web diagrams, resolving the singularity corresponds to
giving a finite length to one of the interior segments, representing a
blowup of a $\mathbb{P}^1$ in the toric geometry. The close relation
between the Riemann surface $\Sigma$ and the toric diagram suggests a
way of reading the effect in the dimer, and hence in the gauge theory,
of the blowup.

The basic idea is that one should identify which external legs of the
web diagram go to which daughter singularity after the resolution. In
the dimer this divides the zig-zag paths into two sets, since we have
a one to one map between zig-zag paths and external legs of the web
diagram of the toric singularity. Let us call the zig-zag paths in the
first set paths of type 1 and those in the second set paths of type
2. In turn, since each edge in the dimer model is crossed by exactly
two zig-zag paths, this divides the set of edges into three, namely
those where two zig-zags of type 1 meet, those were two of type 2
meet, and those where zig-zag paths of mixed type meet. Let us denote
this as edges of type 1, 2 and 3 respectively.

In terms of the mirror surface, a resolution consists of sending a
given set of external legs to infinity (actually finite distance, but
we will be integrating out the corresponding massive mediators). Then
the theory divides into two sectors, one corresponding to each
daughter singularity. In terms of dimer diagrams, what we have is that
the original dimer diagram decomposes into two daughter diagrams, that
we can call 1 and 2. The subdimer 1 is obtained from the original
dimer diagram by removing all edges of type 2, and similarly for the
subdimers of type 2. Edges of type 3 remain in both diagrams. This can
be seen quite intuitively from the mirror, since edges of type 1 and 2
are localized in different sides of the resolution, while edges of
type 3 can communicate with both sectors.

In terms of the gauge theory, what we have done is a Higgsing of the
original theory, where we assign the following vacuum expectation
values to bifundamentals:
\begin{equation}
\Phi_1 = \left(\begin{array}{cc}
0 & 0 \\
0 & v
\end{array}\right),\qquad
\Phi_2 = \left(\begin{array}{cc}
v & 0 \\
0 & 0
\end{array}\right),\qquad
\Phi_3 = \left(\begin{array}{cc}
0 & 0 \\
0 & 0
\end{array}\right),
\end{equation}
where $\Phi_i$ denotes the vev for the fields of type $i$. We see that
these vevs force us to introduce Fayet-Iliopoulos terms in order to
cancel the D-terms and remain in a supersymmetric vacuum. Also, they
trigger the recombination of some gauge factors into their diagonal
combinations, which in terms of the dimer is represented as the
recombination of the two faces adjacent to the edge that gets a vev.

\medskip

\begin{figure}[ht!]
\begin{center}
  \includegraphics[scale=0.7]{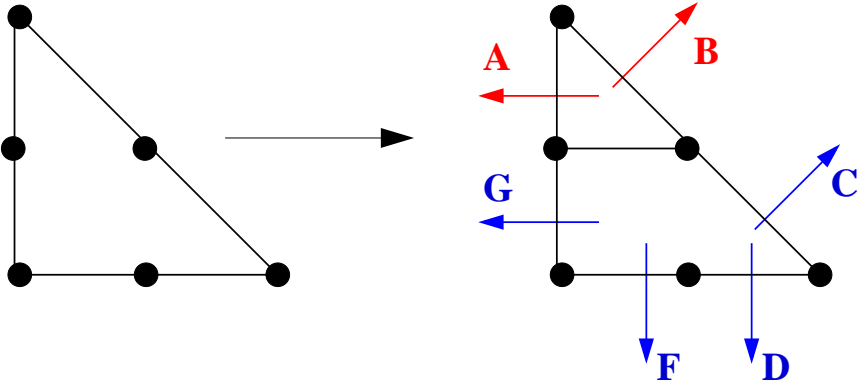}
  \caption{\small $\mathbb{C}^3/\mathbb{Z}_2\times\mathbb{Z}_2 \rightarrow
    SPP$ resolution in terms of the toric diagram. We have also drawn
    superposed, in color, the external legs of the web diagram. The
    desired resolution divides this set of external legs into A,B (in
    red) corresponding to a smooth geometry, and the rest of legs (in
    blue), which describe the SPP daughter singularity.}
  \label{z2z2spptoric}
\end{center}
\end{figure}

Let us show how the procedure works for the
$\mathbb{C}^3/\mathbb{Z}_2\times\mathbb{Z}_2 \rightarrow SPP$
resolution considered in the text \footnote{We do not really need the
  technology described in this section in order to study this
  resolution, simply by giving a vev to any bifundamental of the
  $\mathbb{C}^3/\mathbb{Z}_2\times\mathbb{Z}_2$ theory we end up in
  the SPP theory. We have chosen to describe the general method in
  order to make explicit how the fine details of the discussion in
  Section~\ref{long-general} work. Other more involved examples which
  better illustrate the main idea can be found in
  \cite{GarciaEtxebarria:2006aq}.}. The description of the resolution
we are after in terms of toric diagrams is shown in
Figure~\ref{z2z2spptoric}.

\medskip

\begin{figure}[ht!]
\begin{center}
  \includegraphics[scale=1.3]{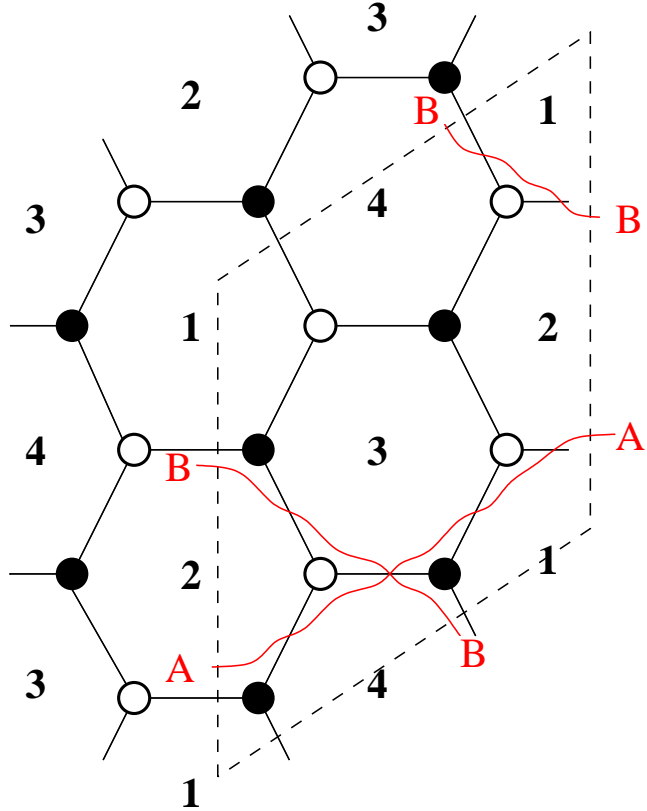}
  \caption{\small The A and B zig-zag paths for the $\mathbb{C}^3/\mathbb{Z}_2\times\mathbb{Z}_2 \rightarrow
    SPP$ orbifold.}
  \label{z2z2zz}
\end{center}
\end{figure}

According to the general procedure outlined in this section, finding
which field gets a vev in this resolution (in order to describe the
SPP side) is just a matter of finding out on which edge A and B
intersect \footnote{Note that there is no invariant way of defining A
  and B in this example, and any other choice of a couple of zig-zag
  paths with different (p,q) charges will lead to the same IR
  theory.}. This edge is localized on the flat space daughter
``singularity'', and disappears from the SPP theory. The relevant
zig-zag paths are shown in Figure~\ref{z2z2zz}. They intersect over
$X_{34}$, and thus this is the field that gets a vev on the SPP side.

 Proceeding on a similar fashion with the C,D zig-zag paths (for
example) one could higgs down the theory to the conifold, as we have
done in the text.

\end{document}